\documentclass[aps, prd, reprint, superscriptaddress, floatfix, showpacs, amsmath, amssymb, hidelinks]{revtex4-1}
\pdfoutput=1

\usepackage{lineno}
\usepackage{graphicx}
\usepackage{dcolumn}
\usepackage{bm}
\usepackage{epstopdf} 
\usepackage[usenames,dvipsnames,pdftex]{color} 
\usepackage{hyperref}
\usepackage[normalem]{ulem} 
\usepackage{subfigure}

\newcommand {\PP}{\textit{p}+\textit{p} }
\newcommand {\EP}{\textit{e}+\textit{p} }
\newcommand {\EA}{\textit{e}+A }
\newcommand {\QTWO}{\ensuremath{Q^2} }
\newcommand {\XB}{\ensuremath{x_{\mathrm{B}}} }
\newcommand {\PT}{\ensuremath{p_{\mathrm{T}}} }
\newcommand {\ALL}{\ensuremath{A_{\mathrm{LL}}} }
\newcommand {\DG}{\ensuremath{\Delta G} }

\begin{document}

\title{Experimental Aspects of Jet Physics at a Future EIC}

\author{B.S.~Page}
\author{Xiaoxuan Chu}
\author{E.C. Aschenauer}
\affiliation{Physics Department, Brookhaven National Laboratory, Upton, NY 11973, U.S.A.}

\date{\today}

\begin{abstract} 
In this work, we present an overview of experimental considerations relevant to the utilization of jets at 
a future Electron-Ion Collider (EIC), a subject which has been largely overlooked up to this point. A comparison of 
jet-finding algorithms and resolution parameters is 
presented along with a detailed analysis of basic jet quantities, such as multiplicities and kinematic distributions. 
A characterization of the energy in the event not associated with a jet is also made. In addition, detector 
requirements and the effects of realistic detector resolutions are discussed. Finally, an example analysis is 
presented in which dijets are used to access the gluon helicity contribution to the spin of the proton.

\end{abstract}

\maketitle

\section{Introduction} \label{sec:INTRO}

Jet observables have proven to be powerful tools for the exploration of the subatomic world in high 
energy collider environments (see for example~\cite{Ali:2010tw}). 
Early jet measurements at $e^{+}e^{-}$ colliders established 
the spin-$1/2$ nature of quarks as well as the existence and spin properties of gluons and helped 
solidify Quantum Chromodynamics (QCD) as the correct theory  of the strong interaction. At 
hadron-hadron colliders (as well as the lepton-proton collider HERA), jets have become indispensable 
tools for studies ranging from the determination of both polarized and unpolarized parton distribution functions (PDFs) to the 
exploration of the electroweak sector to beyond the standard model searches, for examples please see 
\cite{deFlorian:2019zkl,Nocera:2014gqa,Hou:2019efy,Bhatti:2010bf,Marzani:2019hun,Abramowicz:2015mha,Adloff:1997da,Aid:1995du,Derrick:1995bg,Derrick:1996zy} and references therein. Advances in 
background subtraction and substructure techniques have also made jets attractive probes of the 
hot dense medium created in heavy ion collisions \cite{Sadhuon:2019tgz,Connors:2017ptx}. Jet data coming from modern colliders 
such as 
the LHC are being matched by ever more sophisticated theoretical understanding, with calculations 
for many channels reaching next-to-next-to-leading order (NNLO) in $\alpha_s$ and including the 
resummation of relevant large logarithms. The combination of advanced experimental techniques 
and theoretical power make jet observables true precision probes. 

Given the utility of jets in other collider settings, it is logical to explore their potential 
applications at the proposed Electron-Ion Collider (EIC). Several topics important to the 
physics goals of an EIC have been identified which may benefit from jet analyses, including 
accessing the gluon Wigner distribution \cite{Hatta:2016dxp,Hagiwara:2016kam}, probing the linearly polarized 
Weizs\"{a}ecker-Williams gluon transverse 
momentum dependent distributions \cite{Dumitru:2018kuw,Zhou:2016rnt,Boer:2010zf,Pisano:2013cya} and the gluon Sivers function \cite{Zheng:2018awe},
exploring the (un)polarized hadronic structure of the photon \cite{Chu:2017mnm}, 
constraining (un)polarized quark and gluon PDFs at moderate to high momentum fraction ($x$) 
values \cite{Boughezal:2018azh}, and studies of hadronization and cold nuclear matter properties \cite{Liu:2018trl,Arratia:2019vju}. 
There are two features 
inherent to jets which make them attractive probes for the above topics, the first being that jets 
are good surrogates for the scattered partons due to the fact that they contain many of the 
final state particles that arise as the parton hadronizes and thus more accurately represent 
the parton kinematics than single particle observables. The second property is that jests have 
substructure, which characterizes the distribution of energy within the jet in a rigorous way  \cite{Asquith:2018igt,Larkoski:2017jix}. 
Studying how substructure is modified between \EP and \EA collisions could provide information 
about how partons loose energy in the cold nuclear medium \cite{Cao:2017hhk}.

Although several studies of the impact jets may have at an EIC for specific topics have been 
performed, 
no dedicated exploration of the experimental issues surrounding jet finding at an 
EIC has been done. And, while jets were studied extensively at HERA,  
the lower center-of-mass 
energy envisioned for an EIC means that many of the jet properties 
observed at HERA will be different at an EIC. Therefore, this paper systematically details 
several experimental aspects of jet physics as they are expected to manifest at an EIC, as 
well as outlining an example analysis. We hope this paper will serves as a resource for those 
interested in exploring the use of jets at a future EIC for a wide range of physics topics.

The remainder of the paper is organized as follows: Sec. \ref{sec:SIMU} describes the Monte Carlo 
setup used to generate the \EP events we analyze. In Sec. \ref{sec:JETFINDING}, we discusses 
several aspects of jet finding, such as the choice of algorithm and recombination parameter as 
well as some basic kinematic properties of jets at an EIC. Section \ref{sec:UE} characterizes 
energy and particle distributions from the underlying event, 
as provided by our Monte Carlo, which will add additional energy to the reconstructed jets. 
In addition the simulated underlying event activity is compared to similar measures from \PP 
events at $\sqrt{s} = 200$~GeV performed by the STAR experiment. The effect that a realistic 
detector will have on the energy resolution of reconstructed jets is explored in 
Sec. \ref{sec:DETECTOR} by smearing the momentum and energy of incoming particles according 
to a specific detector model. Special attention is given to the role of hadronic calorimetry at 
mid-rapidity. Section \ref{sec:DELTAG} presents a method for accessing gluon polarization by 
measuring the longitudinal double-spin asymmetry for dijet final states. Strategies for 
isolating the appropriate partonic subprocesses utilizing cuts on the dijet kinematics are 
discussed and the value and statistical precision of the expected asymmetries are shown and 
compared with our current knowledge of the polarized parton distribution functions. 
Finally, we summarize our findings in section \ref{sec:SUMMARY}.

\section{Monte Carlo Sample} \label{sec:SIMU}

To facilitate the studies presented in this paper, a large sample of \EP pseudo-data 
was generated using PYTHIA-6.4 \cite{Sjostrand:2006za}. 
Two proton PDFs were 
used as input, depending on the \QTWO range being simulated. For \QTWO greater than 
1.0~GeV$^2$, the CTEQ6.1 PDF \cite{Stump:2003yu} set was used, while the CTEQ5m set \cite{Lai:1999wy} was used for 
$10^{-5}<Q^{2}<1.0$~GeV$^2$. The SAS PDF set \cite{Schuler:1995fk,Schuler:1996fc} was used for cases when the partonic structure of the photon was relevant. The choice of the lower \QTWO limit was driven by the acceptance of a proposed low $Q^2$-tagger, which will reside near the 
beampipe, outside of the central detector.
The CTEQ5m PDF is used for the proton, because contrary to modern PDFs (i.e., CT, NNPDF, 
HERAPDF, MSTW) its value is not frozen at its input scale $Q^2_0$, but allows description 
of the partonic structure of the proton at $Q^2 \leq Q^2_0$.
In addition, this set reproduces the low \QTWO HERA cross sections quite well 
(see Figure 2 in \cite{Chu:2017mnm}).

Events were generated at center-of-mass energies $(\sqrt{s})$ of 45~GeV and 141~GeV, corresponding to 
(electron x proton) beam energies of 10 x 50~GeV and 20 x 250~GeV, respectively. 
These energies represent lower and higher ranges generally considered for an EIC \cite{Accardi:2012qut}. As of 
this writing, EIC machine designs \cite{Aschenauer:2014cki} are still being finalized, so the ultimate maximum and minimum 
beam energies may deviate somewhat from those assumed in this paper, but should not greatly affect the 
conclusions drawn here. It is also envisioned that the 
EIC will be able to operate at a number of $\sqrt{s}$ values between these bounds. 

\begin{figure*}
	\begin{center}
		\subfigure[]{
			\centering
			\includegraphics[width=0.25\textwidth]{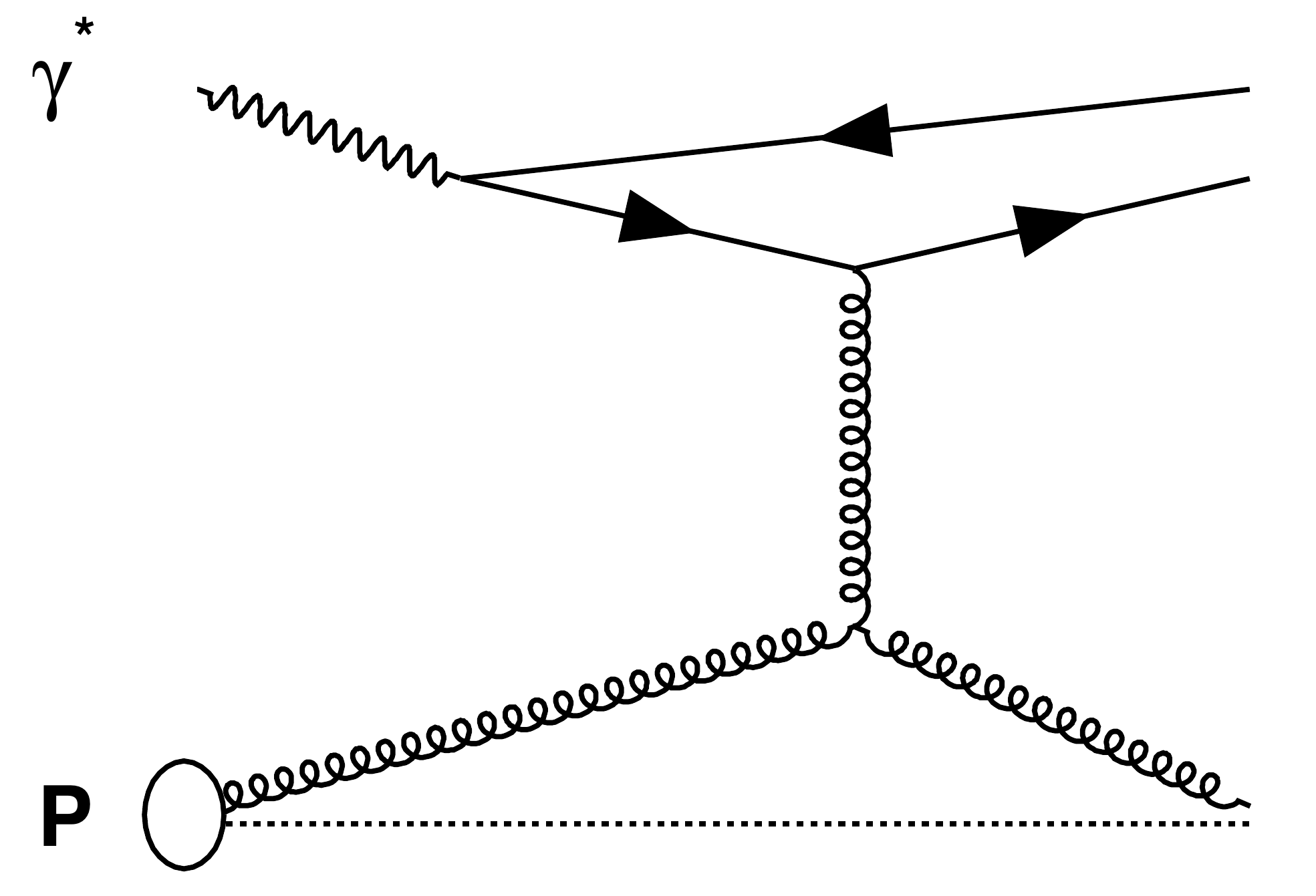}
			\label{fig:RESOLVED}
		}
		\quad
		\subfigure[]{
			\centering
			\includegraphics[width=0.25\textwidth]{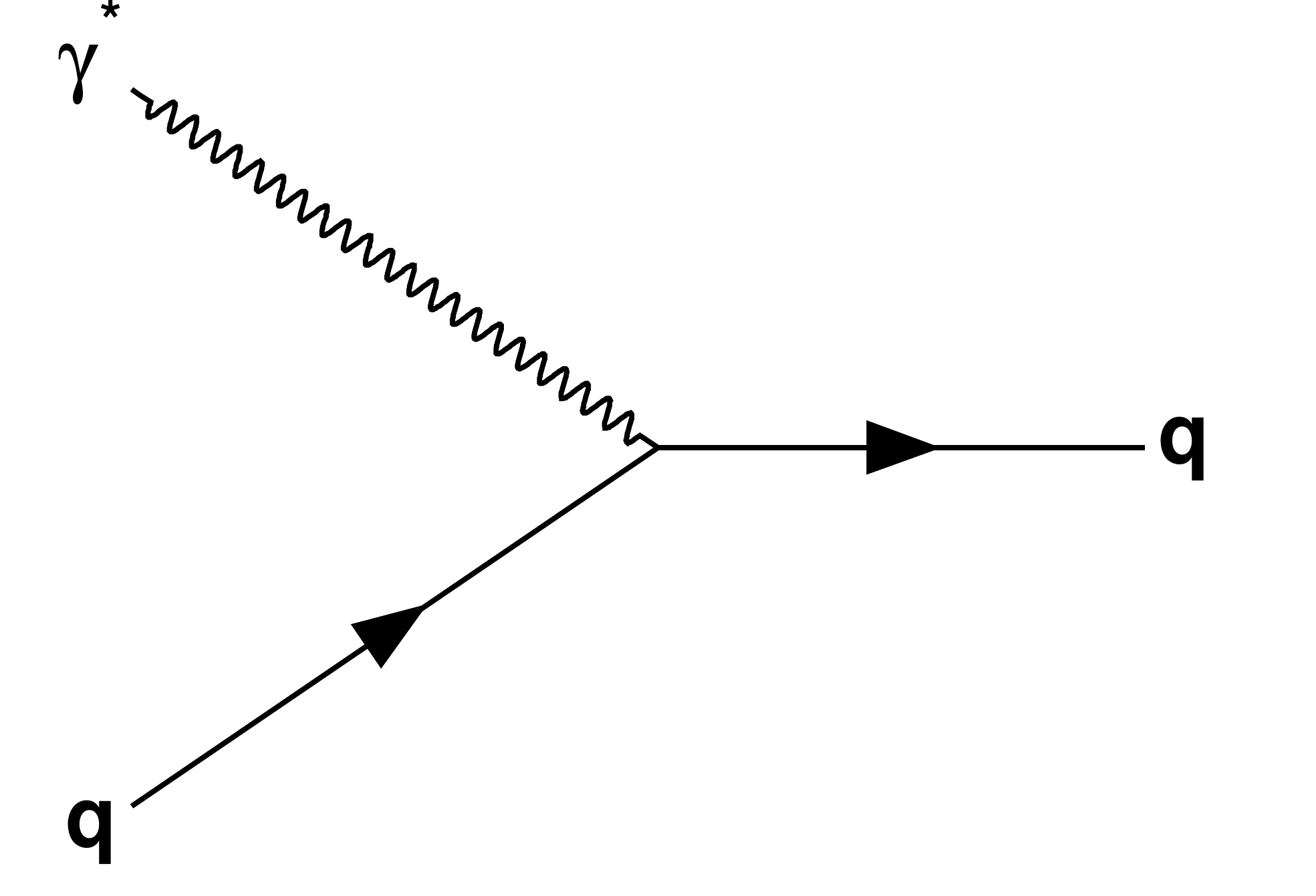}
			\label{fig:LODIS}
		}
		
		\subfigure[]{
			\centering
			\includegraphics[width=0.25\textwidth]{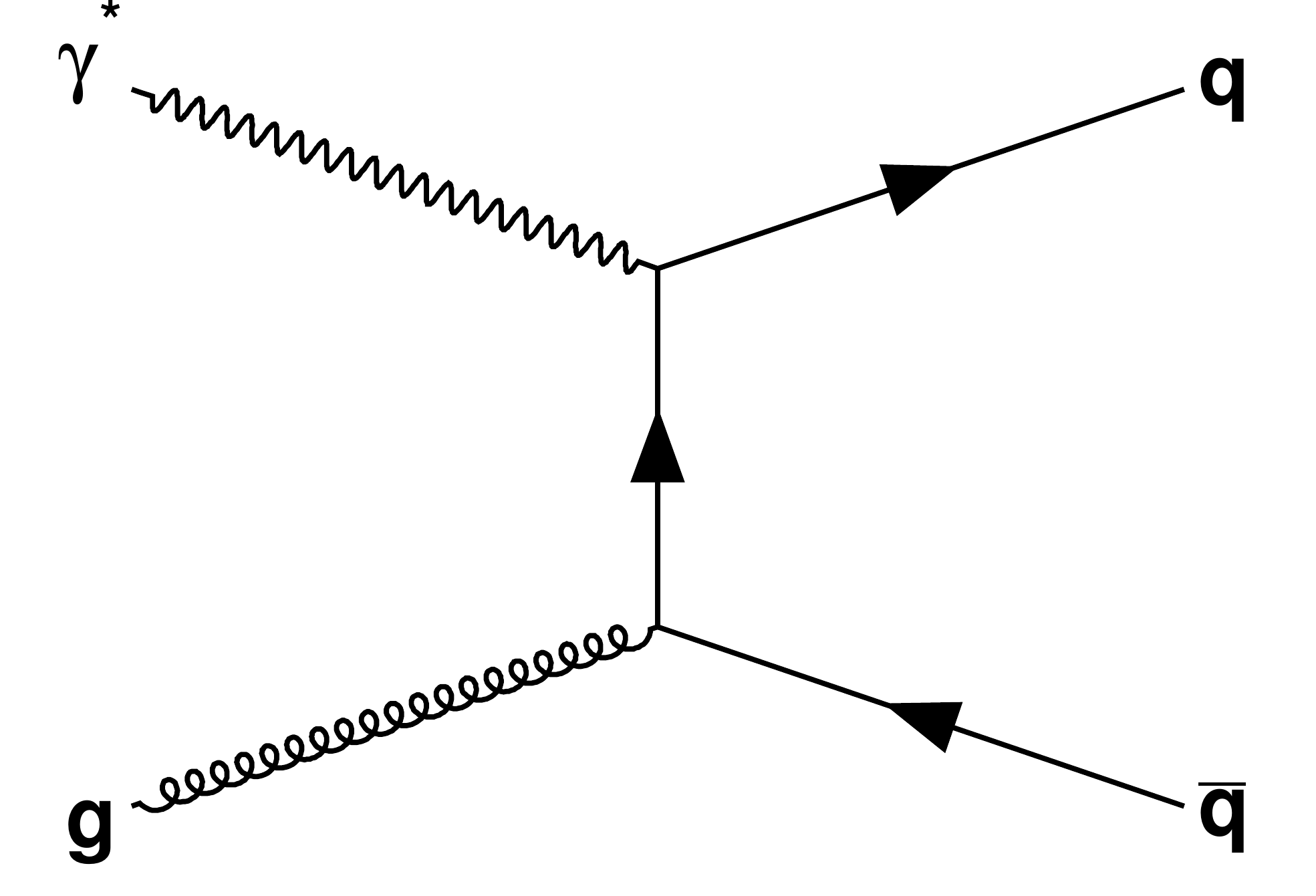}
			\label{fig:PGF}
		}
		\quad
		\subfigure[]{
			\centering
			\includegraphics[width=0.25\textwidth]{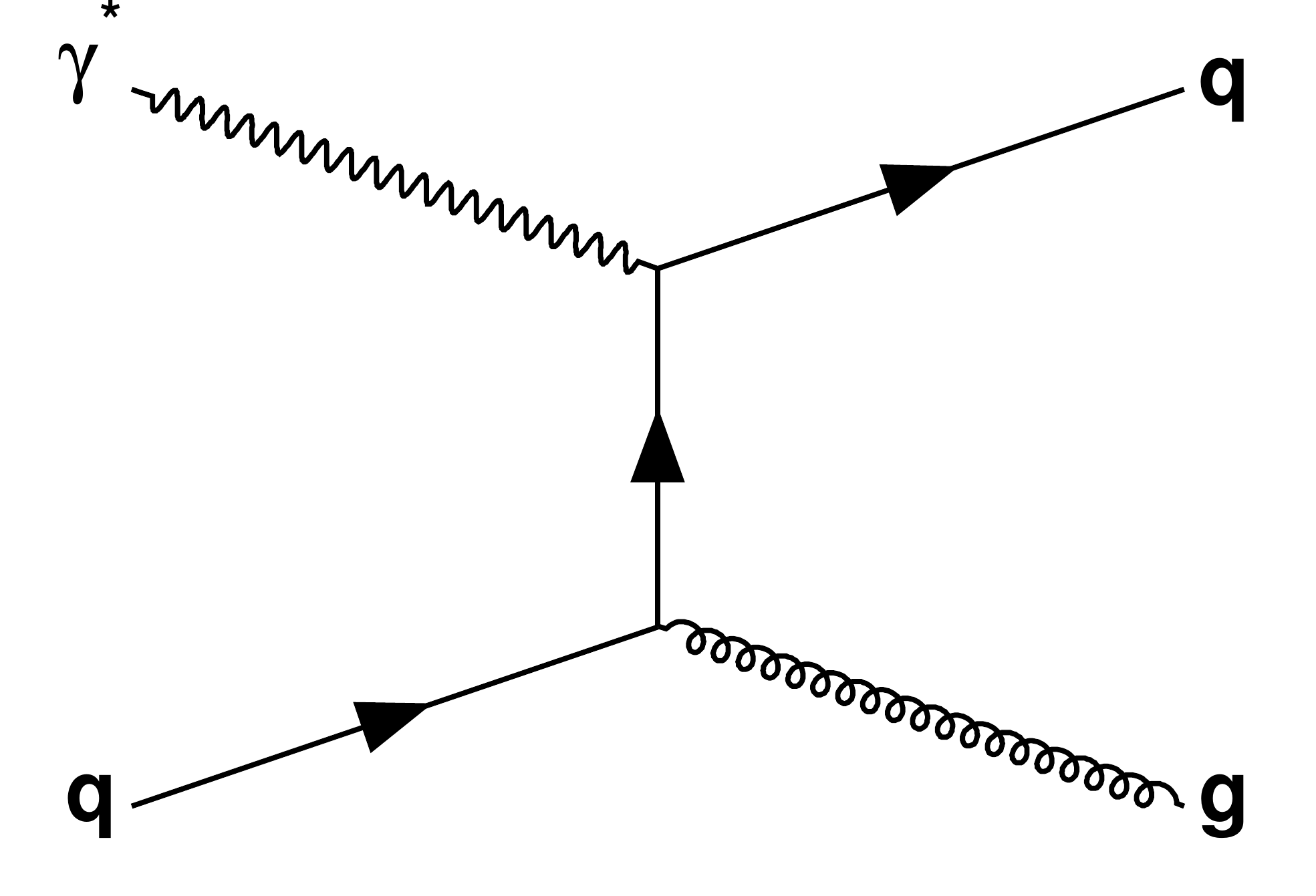}
			\label{fig:QCDC}
		}
		\caption[direct subprocess]{Feynman diagrams for the subprocesses considered in this analysis: 
               (a) Resolved (b) $\mathcal{O}(\alpha^{0}_{s}) $ LO DIS, (c) Photon-Gluon Fusion (PGF) 
               and (d) QCD Compton scattering (QCDC). Figure is from~\cite{Zheng:2014vka}.}
		\label{fig:DISgraph}
	\end{center}
\end{figure*}

In this paper, we only consider neutral current (NC) events, a first study on the capabilities 
of an EIC for charged current (CC) events can be found in \cite{Aschenauer:2013iia}. The NC events fall into two 
categories: resolved and direct.
Resolved processes (see Fig. \ref{fig:RESOLVED}) are those in which the virtual 
photon interacts via the hadronic component of its wavefunction, contributing a quark or gluon 
to a hard-scattering with a parton from the nucleon. The resolved category includes the 
$qq\rightarrow qq, \, q\bar q \rightarrow q \bar q, \, q\bar q\rightarrow gg, \,
qg\rightarrow qg, \, gg\rightarrow q\bar q, gg\rightarrow gg$ subprocesses and plays a 
significant role in the production of high-\PT particles at low \QTWO. 
In direct processes, the photon interacts as a point-like particle with the partons of the 
nucleon. The subprocess which comprise the direct category include leading order 
DIS \ref{fig:LODIS} (L.O. subprocess), photon-gluon fusion (PGF) \ref{fig:PGF}, and QCD Compton 
(QCDC) \ref{fig:QCDC}. Direct processes contribute more at high \QTWO values while resolved 
processes dominate at $\QTWO < 1$~GeV$^2$. The higher order resolved, QCDC, and PGF subprocesses (hereafter referred to as the H.O. subprocesses) involve two 
high-\PT partons separated in azimuth and thus often give rise to dijet events.

\section{Jet Finding and Jet Properties} \label{sec:JETFINDING}

The jets used in this study were formed using stable final state particles generated
by the PYTHIA Monte Carlo described in section \ref{sec:SIMU}. Here, stable refers 
to particles which would not normally decay in the volume of a detector, 
such as charged pions, kaons, protons, and neutrons. Neutral pions were allowed to decay and 
the resulting (predominately) photons were passed to the jet finder. 
To match expected detector acceptances, only particles having transverse momenta with respect 
to the beam greater than 250~MeV/$c$ 
and pseudorapidity between $\pm 4.0$ 
were considered candidates for 
inclusion in jets; the scattered lepton was not allowed to be part of the jet. The effects of 
using a higher minimum particle \PT cut were also explored (see Sec. \ref{sec:MINPARTPT}).

\subsection{Reference Frames} \label{sec:REFFRAMES}
At hadron colliders, analyses are carried out almost exclusively in the reference frame 
of the detector, the laboratory frame, as the kinematics of the interacting partons are 
not generally known. For DIS, however, the scattering kinematics are known 
event-by-event which makes it possible to boost to other frames. A particularly useful 
frame for jet analyses is the Breit or `brick wall' frame \cite{Rinehimer:2009yv}. The Breit frame is oriented 
such that for the lowest order DIS process $\gamma^{*}q \rightarrow q^{\prime}$, the 
virtual photon and interacting quark collide head-on along the z-axis and is boosted such 
that the only non-zero component of the virtual photon four-momentum  is $p_z = -Q$. A 
consequence of this boost is that the z-momentum of the incoming quark is $Q/2$ while 
the scattered quark has z-momentum $-Q/2$ (hence the name `brick wall' frame) and the 
proton remnant has a z-momentum of $(1-x)Q/(2x)$. This leads to a natural separation 
between jets associated with the struck quark and those associated with the proton 
remnant. Working in the Breit frame has the effect of suppressing contributions from the 
L.O. subprocess, as the scattered quark has zero transverse momentum by construction (although, as will be seen 
in \ref{sec:INCKIN}, jets can arise from the L.O. process at large \QTWO due to final state radiation). 
A dedicated study of L.O. jets in the laboratory frame can be found in \cite{Arratia:2019vju}.
The Breit frame is also advantageous for the study of jets arising from the H.O. subprocesses as their 
transverse momenta will be taken with respect to the photon-quark axis, which is the relevant quantity in these 
interactions. This is especially important at large values of \QTWO where the angle 
between the photon-quark and beam axes is significant. Boosts to other frames such as the photon-quark 
center-of-mass or proton rest frames can also be performed, but are not covered here.

For this analysis, all particle four-momentum were boosted into the Breit frame and then 
passed to the jet finder for clustering. This avoids any changes in the particle content 
which may arise due to variations in clustering between the two frames. When it is necessary to 
present a jet quantity with respect to the detector, the Breit frame thrust axis of the jet 
was simply boosted back to the laboratory frame.

\subsection{Jet Definitions} \label{sec:JETDEFS}

While the idea of a jet as a collimated spray of particles is conceptually easy to grasp 
and jets are often easy to identify `by eye' in event displays, a well defined 
method of mapping a set of particles into a set of jets is required for jets to be useful 
in experimental and theoretical analyses.
The collection of rules that determine how particles are grouped into jets is 
known as a jet algorithm, while the prescription for merging the momenta of individual 
particles to form the overall jet momentum is known as a recombination scheme. The 
combination of jet algorithm, recombination scheme, and any additional parameters 
controlling the behavior of the jet algorithm is known as a jet definition and, as the 
name implies, fully defines a jet for purposes of an analysis \cite{Salam:2009jx}.

There are a number of jet algorithms on the market \cite{Salam:2009jx,Ellis:2007ib} and the determination of which algorithm to 
utilize will depend on the requirements of the specific analysis being performed. As this manuscript 
is meant to give a general overview of jets at an EIC, the choice of a specific algorithm is not 
critical as long as there are not significant differences between algorithms for basic jet 
properties. To confirm this is the case, a number of jet quantities including yields, particle content, 
energy profile, transverse momenta, and rapidities were compared using the anti-k$_T$ \cite{Cacciari:2008gp}, 
k$_T$ \cite{Ellis:1993tq}, and SISCone \cite{Salam:2007xv} algorithms. 
These were chosen as they represent the two broad categories of algorithms, 
sequential recombination and cone, and have seen use at both HERA and hadron colliders such as the LHC.
Differences in the studied quantities ranged from negligible to a couple of percent, 
indicating that the choice of algorithm will have little impact on the conclusions drawn here. Because 
it produces jets with regular boundaries that are slightly more collimated, 
the anti-k$_T$ algorithm will 
be used for all subsequent studies in this paper. It should be noted that several recombination schemes 
also exist, but only the E\_Scheme \cite{Blazey:2000qt}, in which particles are combined simply by adding their 
four-momenta, will be considered here.


The other major component of the jet definition which 
needs to be determined is the resolution parameter, $R$, which sets the effective size of the jet. 
As with jet algorithm, the 
optimal choice for $R$ will depend on requirements driven by a particular analysis. 
Often, the chosen $R$ is a compromise between large values that will capture more energy from 
the hadronizing partons and small values, which limit the contamination from underlying event (see 
Sec. \ref{sec:UE} for a discussion of expected underlying event activity at an EIC). 
The resolution parameter can be expected to affect the jet yield and particle content of jets, 
and because it influences how much of the energy from the hadronizing parton ends up in the jet, $R$ 
should also affect how well the jet represents the underlying partonic behavior.

\begin{figure}[t!]
	\centering \includegraphics[keepaspectratio=true, width=0.95\linewidth]{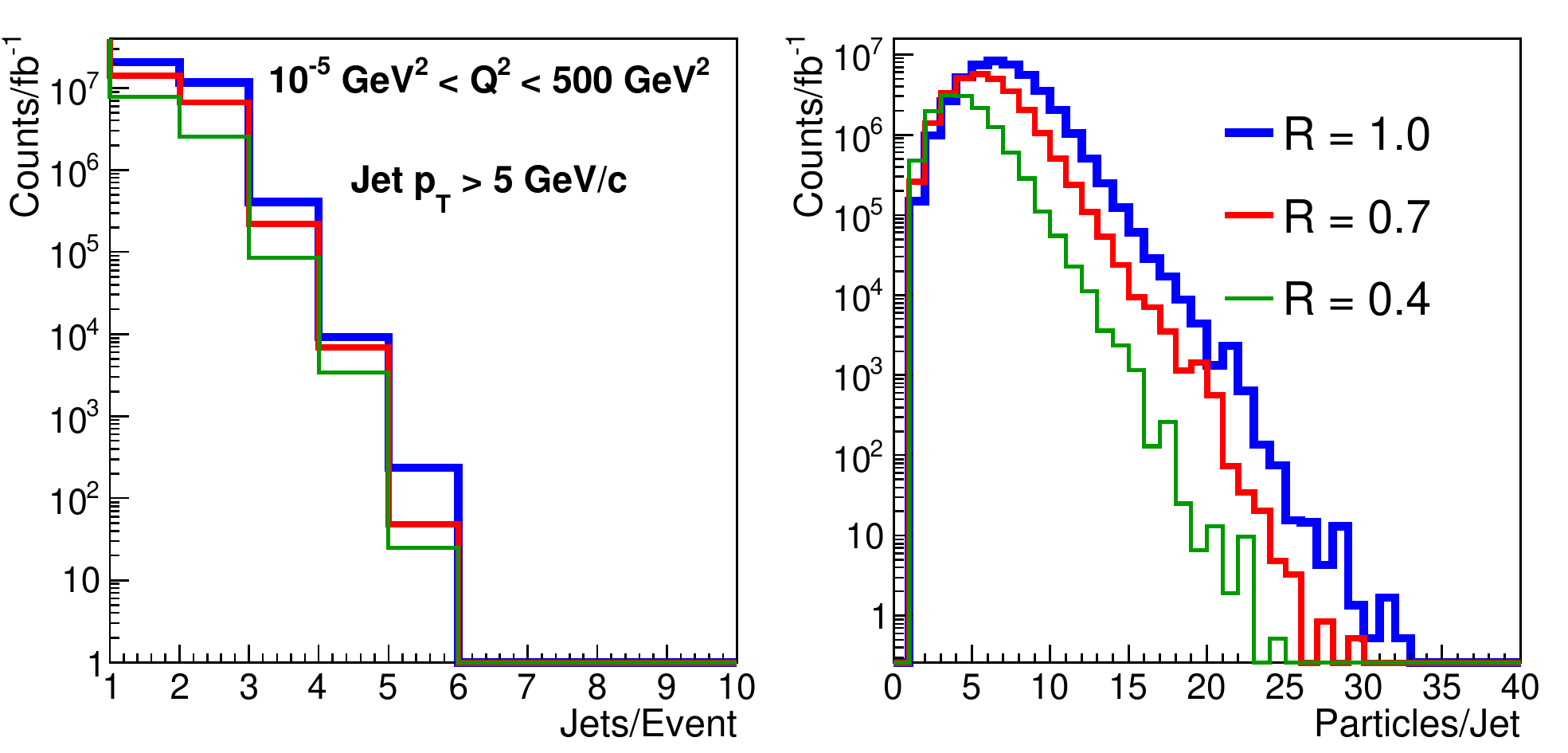}
	\caption[Radius Comparison Multiplicity]{[color online] Comparison of jet multiplicity (left panel) and particle multiplicity
    within the jet (right panel) for the anti-k$_{T}$ algorithms and three resolution parameters $R$ = 1.0, 0.7, 
    and 0.4. The \QTWO range is between $10^{-5}$~GeV$^{2}$ and 500~GeV$^{2}$ and the Resolved, QCDC, PGF, and leading order DIS subprocesses have been combined.}
	\label{fig:RADMULTCOMP}
\end{figure}

The jet multiplicity and number of particles within a jet are presented in 
the left and right panels, respectively, of Fig. \ref{fig:RADMULTCOMP} for 
$R$ values of 1.0, 0.7, and 0.4. The jets were required to have 
transverse momenta greater than 5~GeV/$c$ in the Breit frame and were taken from events with 
$Q^2$ between $10^{-5}$ and 500~GeV$^2$. The L.O. and H.O. subprocesses have been 
combined. It is seen that the total number of found jets 
increases with increasing $R$, which is due to the fact that larger $R$ values admit more particles 
(as seen in the right panel), meaning more jets will pass the 5~GeV/$c$ transverse momentum threshold.

\begin{figure*}[t!]
	\centering \includegraphics[keepaspectratio=true, width=0.95\linewidth]{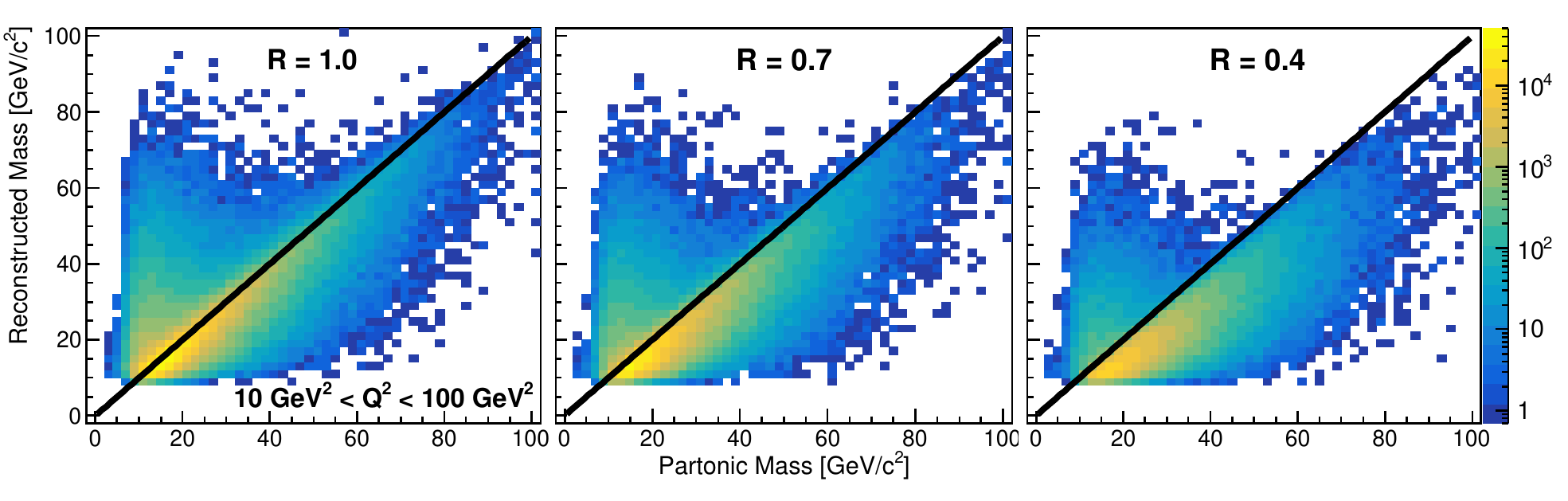}
	\caption[Radius Comparison Mass]{[color online] Dijet invariant mass compared to the partonic invariant mass
		$\sqrt{\hat{s}}$ for $R =$ 1.0 (left), 0.7 (middle), and 0.4 (right). Event \QTWO was required to between 10~GeV$^{2}$ and 100~GeV$^{2}$ and the Resolved, QCDC, PGF, and leading order DIS subprocesses have been combined.}
	\label{fig:RADMASSCOMP}
\end{figure*}

One benefit of jet observables is that they can serve as proxies for hard-scattered partons 
because the 
jet will capture many of the particles emitted as the parton hadronizes. Since the size of $R$ will 
affect the amount of partonic radiation included in the jet, it is necessary to assesses how changes 
in $R$ impact how well jets reproduce the underlying partonic kinematics. This was done by comparing the 
reconstructed dijet invariant mass to the parton level invariant mass.
Dijets were reconstructed from H.O. 
events by selecting the two jets with the highest transverse momentum in the Breit frame, 
requiring that they be greater than 120 degrees apart in azimuth, and necessitating that one jet have \PT greater 
than 5~GeV/$c$ while the other has \PT greater than 4~GeV/$c$. The comparison between reconstructed dijet 
and partonic invariant mass can be seen in Fig. \ref{fig:RADMASSCOMP} for $R = $ 1.0, 0.7, and 0.4. Only \QTWO 
values between 10~GeV$^{2}$ and 100~GeV$^{2}$ are shown as the conclusions are the same for all \QTWO values.
The best agreement is seen for the largest $R$ value and degrades as $R$ decreases. The events at low 
partonic and large reconstructed dijet invariant mass arise when the one or both of the jets comprising the dijet do not 
match the true outgoing partons. With the exception of these `false' dijets, the fact that 
the reconstructed mass is consistently smaller than the partonic mass for $R < 1.0$ indicates 
that the smaller cones are not capturing the full energy associated with the hard-scattered partons.

\begin{figure}[t!]
 \centering \includegraphics[keepaspectratio=true, width=0.75\linewidth]{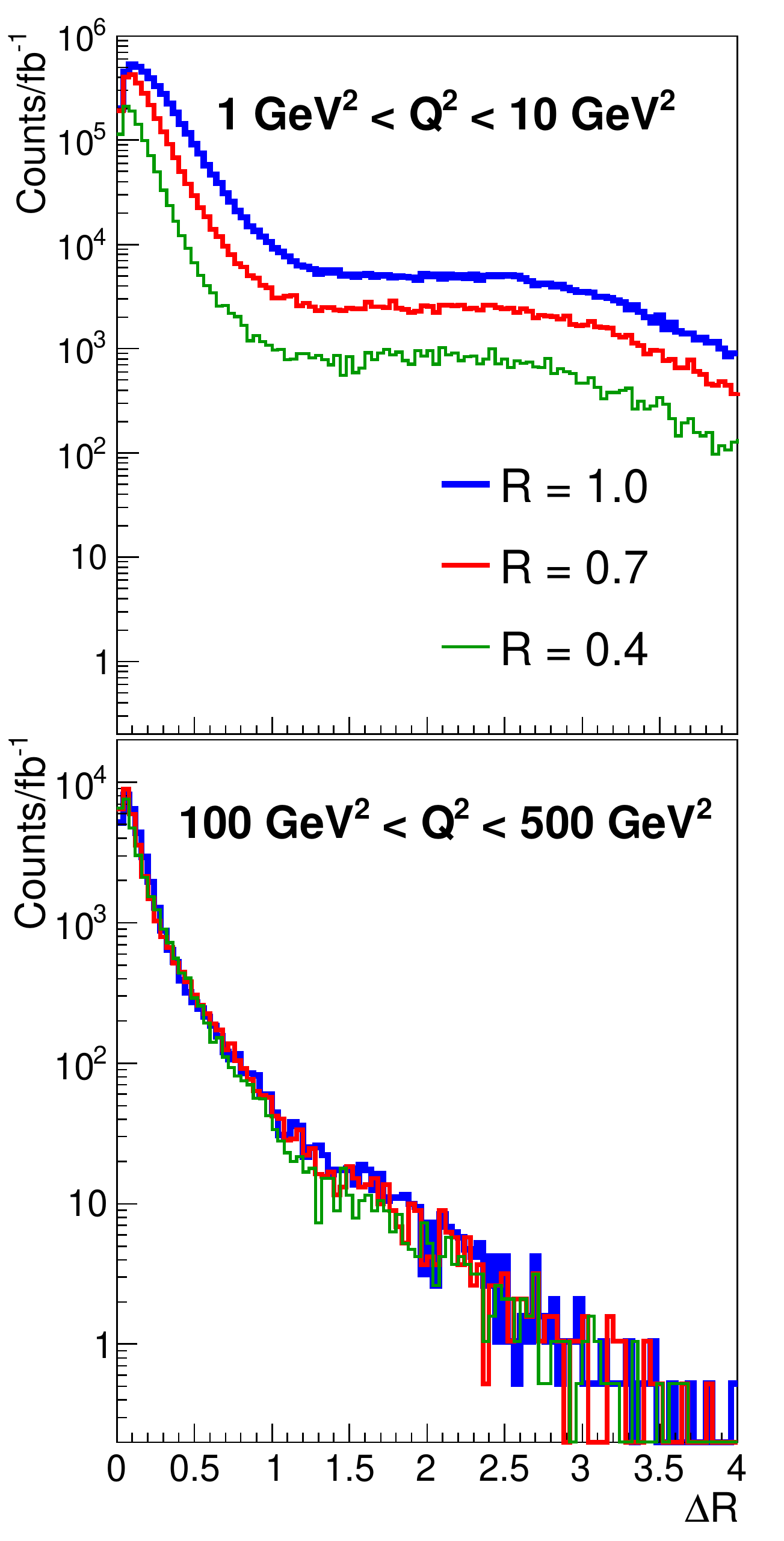}
 \caption[Jet-Parton Delta R]{[color online] Distance in rapidity-azimuth space between the jets comprising a dijet and the corresponding hard-scattered partons with \QTWO between 1~GeV$^{2}$ and 10~GeV$^{2}$ (upper panel) and 100~GeV$^{2}$ and 500~GeV$^{2}$ (lower panel) for jet radii of 1.0, 0.7, and 0.4.}
	\label{fig:JETDELTAR}
\end{figure}

Another way to assess how well jets represent the underlying partons is to measure the degree to which the 
jet thrust axes correspond to the directions of the partons which give rise to them. This is quantified using 
the distance measure $\Delta R$, which is the quadrature sum of the difference between jet and parton rapidity and 
azimuthal angle $(\Delta R \equiv \sqrt{\Delta \eta^2 + \Delta \phi^2})$. For each jet comprising the dijet, 
$\Delta R$ between that jet and the two hard scattered partons is found and the minimum is taken. Figure 
\ref{fig:JETDELTAR} presents this minimum $\Delta R$ for both jets from the H.O. subprocesses 
combined for two \QTWO bins. The $\Delta R$ distributions are larger at low \QTWO and peak closer to zero as 
\QTWO increases. For \QTWO between 1 and 10~GeV$^2$, over 85\% of the jets are within a $\Delta R$ of 0.5 to 
their matching parton and this jumps to 95\% for \QTWO between 100 and 500~GeV$^2$. Again, the tails at large 
$\Delta R$ arise when one or both jets in the dijet do not correspond to one of the true hard scattered partons. 
Surprisingly, at low \QTWO, the jet-to-parton matching is seen to improve as $R$ decreases. This may be due to the 
effect of underlying event activity which will be more prevalent in larger cones and could pull the jet thrust axis 
slightly away from the parton direction, or it may be a selection bias effect wherein jets found with a smaller 
$R$ tend to fragment in a more collimated way. Regardless, Fig. \ref{fig:RADMASSCOMP} shows that any benefit from 
the slightly better matching between the jet and parton directions at small $R$ is overwhelmed by the parton energy 
missed by the smaller jet cone.

As it results in the most jets found with the largest particle content and best reproduces the 
partonic kinematics, $R$ will be set to 1.0 for all subsequent studies presented in this paper. 
Together with the choice of anti-k$_T$ for the jet algorithm and E\_Scheme recombination, 
the jet definition is fully quantified.

\subsection{Inclusive Jet Kinematics} \label{sec:INCKIN}

While jets arising from \EP collisions were studied extensively at HERA, 
the lower center-of-mass energies (maximum $\sqrt{s}$ of 141~GeV for EIC compared to 320~GeV for HERA) 
envisioned for an EIC make a detailed investigation of jet properties 
and kinematics warranted. Jets were found using the jet definition outlined in section 
\ref{sec:JETDEFS} as implemented in FastJet-3.3.1 \cite{Cacciari:2011ma} and were required to have transverse 
momenta greater than 5~GeV/$c$ in the Breit frame. 

\begin{figure}[t!]
	\centering \includegraphics[keepaspectratio=true, width=0.95\linewidth]{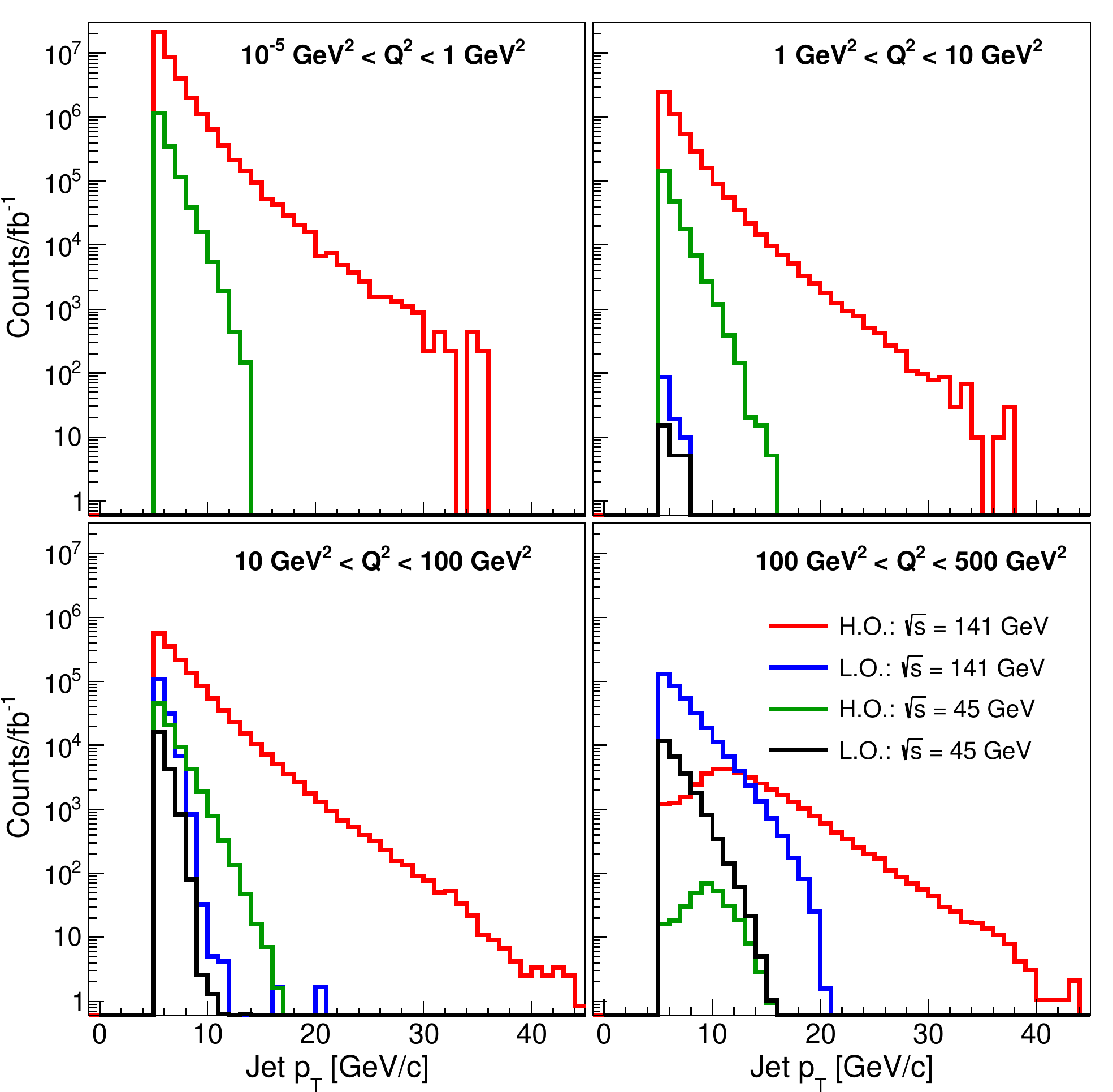}
	\caption[Inclusive Jet \PT Spectra]{[color online] Breit frame inclusive jet \PT spectra for $\sqrt{s} = 141$ and 45~GeV in \QTWO bins of $10^{-5}-1.0$~GeV$^2$ (upper left), 1-10~GeV$^2$ (upper right), 10-100~GeV$^2$ (bottom left), and 100-500~GeV$^2$ (bottom right). The Resolved, QCDC, and PGF subprocesses have been combined and are compared to the leading order DIS spectra and the histograms have been scaled to the counts expected for an integrated luminosity of 1~fb$^{-1}$.} 
	\label{fig:JETPTSPECTRA}
\end{figure}

Breit frame inclusive jet \PT spectra are shown in Fig. \ref{fig:JETPTSPECTRA} for four \QTWO ranges between 
$10^{-5}$ and 500~GeV$^2$ and for $\sqrt{s}$ values of 45 and 141~GeV. The spectra have been 
scaled to the number of counts expected for 1~fb$^{-1}$ of integrated luminosity. As their behaviors 
are similar, the H.O. subprocesses have been combined and are compared with the 
L.O. spectra. It is seen that high-\PT jet production is dominated by the H.O. subprocesses while the L.O. spectra are much softer. 
This is expected because at leading order in the Breit frame, the scattered parton moves along the z-axis and transverse 
momentum is generated primarily via final state radiation. The prevalence of this radiation 
decreases with decreasing $Q^2$, so leading order DIS jets are basically absent for \QTWO below 
10~GeV$^2$. Note that the L.O. jet yield can increase up to two orders of magnitude for $\QTWO > 1$~GeV$^2$ when jet-finding is performed in the 
lab frame, meaning this frame will be preferred for measurements focusing on L.O. jet production \cite{Arratia:2019vju}. 
Figure \ref{fig:JETPTSPECTRA} also demonstrates the critical importance of higher 
center-of-mass energies for jet studies as the yields can be orders of 
magnitude larger for $\sqrt{s} = 141$~GeV compared to $\sqrt{s} = 45$~GeV. This is most pronounced at 
large \PT where yield differences are so great, no practical increase in luminosity could compensate.

\begin{figure*}[t!]
 	\centering \includegraphics[keepaspectratio=true, width=0.95\linewidth]{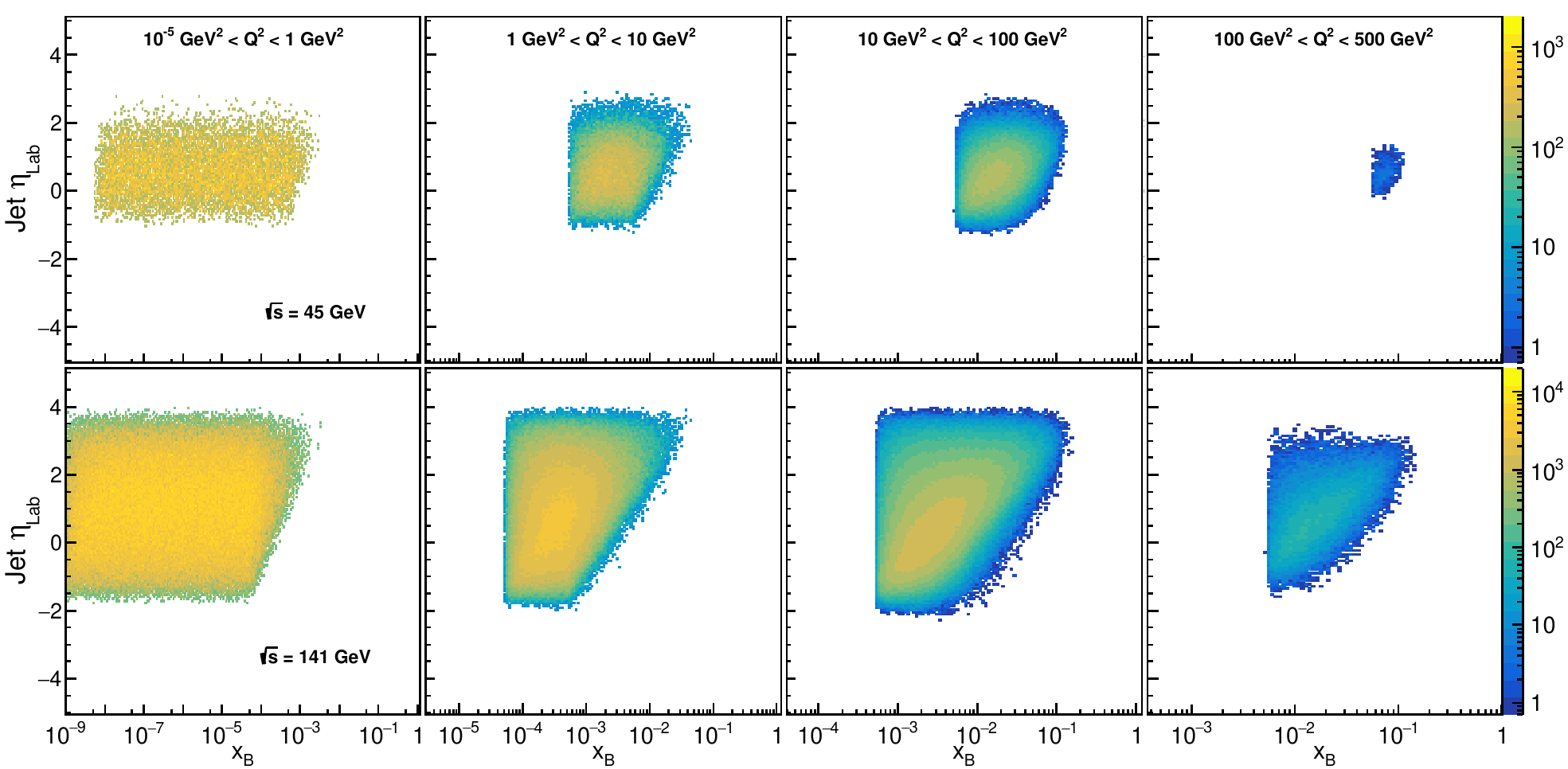}
 	\caption[Inclusive Jet Laboratory Pseudorapidity (RQP)]{[color online] Inclusive jet laboratory pseudorapidity vs \XB for 
 	\QTWO bins of $10^{-5} - 1.0$~GeV$^2$ (left column), 1-10~GeV$^{2}$ (middle-left column), 10-100~GeV$^{2}$ (middle-right column), and 100-500~GeV$^{2}$ (right column) for center-of-mass energies of 45~GeV (upper row) and 141~GeV (bottom row). The resolved, QCDC, and PGF subprocesses are shown. Note that the top and bottom rows are separately scaled to the counts expected for 1~fb$^{-1}$ of integrated luminosity.} 
 	\label{fig:JETETARQP}
\end{figure*}

\begin{figure}[t!]
	\centering \includegraphics[keepaspectratio=true, width=0.95\linewidth]{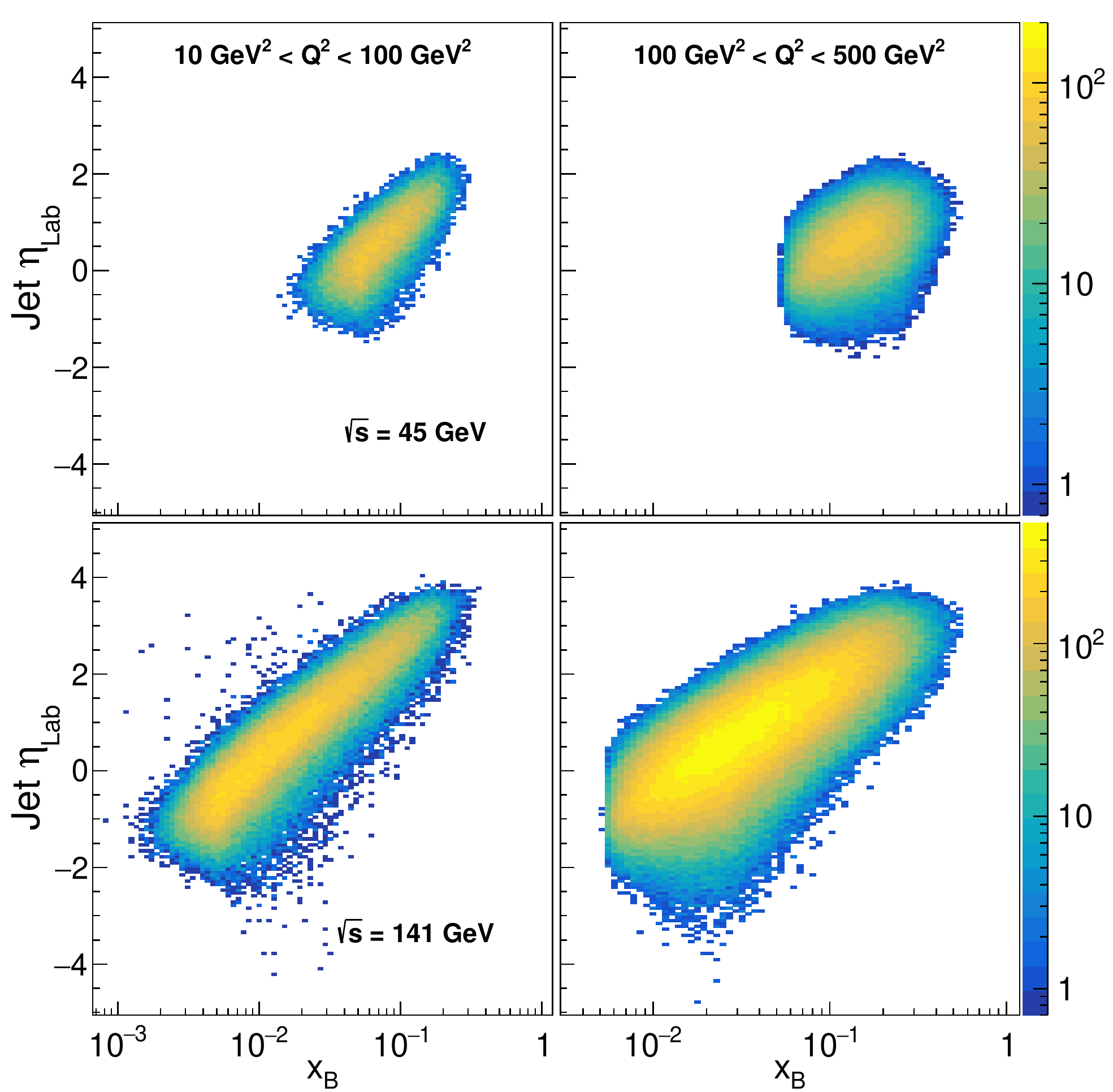}
	\caption[Inclusive Jet Laboratory Pseudorapidity (DIS)]{[color online] Inclusive jet laboratory pseudorapidity vs \XB for \QTWO bins of 10-100~GeV$^{2}$ (left column) and 100-500~GeV$^{2}$ (right column) and center-of-mass energies of 45~GeV (upper row) and 141~GeV (bottom row). Only the leading order DIS subprocess is shown. Note that the top and bottom rows are separately scaled to the counts expected for 1~fb$^{-1}$ of integrated luminosity.} 
	\label{fig:JETETADIS}
\end{figure}

In addition to their typical transverse momenta, it is important to understand where jets 
are located in the detector and how that correlates to the \XB and \QTWO of the event. Figure 
\ref{fig:JETETARQP} shows the inclusive jet pseudorapidity distributions, in 
the laboratory frame, as a function of \XB for four \QTWO bins ranging from $10^{-5}$~GeV$^2$ to 500~GeV$^2$ 
and $\sqrt{s}$ values of 45~GeV and 141~GeV, for the H.O. subprocesses. Figure 
\ref{fig:JETETADIS} shows the same for the L.O. subprocess in the \QTWO ranges 10-100~GeV$^2$ and 
100-500~GeV$^2$. As expected from basic DIS kinematics, smaller 
\XB values are probed by larger center-of-mass energies and visa versa. Comparing Figs. \ref{fig:JETETARQP} and 
\ref{fig:JETETADIS}, it is seen that the leading order DIS jet pseudorapidity is more strongly 
correlated with \XB then that of the Resolved, QCDC, or PGF jets. Because there is only one outgoing parton 
in DIS, jet pseudorapidity should be strongly determined by the event \XB and \QTWO with the observed 
width of the distributions in Fig. \ref{fig:JETETADIS} due to the finite \QTWO ranges and, more importantly, 
final state radiation altering the trajectory of the outgoing quark. The presence of a second hard parton 
in H.O. events breaks the strong relationship between $x_\mathrm{B}$, $Q^2$, and $\eta$ and allows 
the resultant jets to fill the kinematically allowed phase space. The importance of hadron beam energy 
to jet position can also be seen in Figs. \ref{fig:JETETARQP} and \ref{fig:JETETADIS} by contrasting the 
distributions at the two $\sqrt{s}$ values for given \XB and $Q^{2}$. Larger hadron beam momenta impart more 
of a boost to final state particles, so jets at $\sqrt{s} = 141$~GeV will be pushed to higher 
pseudorapidities compared to jets from collisions at lower $\sqrt{s}$. Thus, good forward tracking and 
calorimetry capabilities will be needed to utilize jets at large $\sqrt{s}$. 

\subsection{Dijet Kinematics} \label{sec:DIKIN}

So far, only inclusive jet quantities have been considered, yet as stated above, the H.O. 
subprocesses naturally give rise to correlated two jet final states (dijets). By measuring 
the properties of both jets in coincidence, dijets can provide information on the leading order 
kinematics of the hard scattering event, such as the momentum fraction contributed by the virtual 
photon. Several studies have already explored the utility of dijet measurements at the EIC \cite{Chu:2017mnm,Zheng:2018awe,Dumitru:2018kuw} 
and a further study will be presented in Sec. \ref{sec:DELTAG}. As before, dijets were selected by 
identifying the two jets with the largest transverse momenta in 
the Breit frame and requiring them to be greater than 120 degrees apart in azimuth. It was further 
required that one jet have \PT greater than 5~GeV/$c$ while the other have \PT greater than 
4~GeV/$c$.

\begin{figure}[t!]
	\centering \includegraphics[keepaspectratio=true, width=0.95\linewidth]{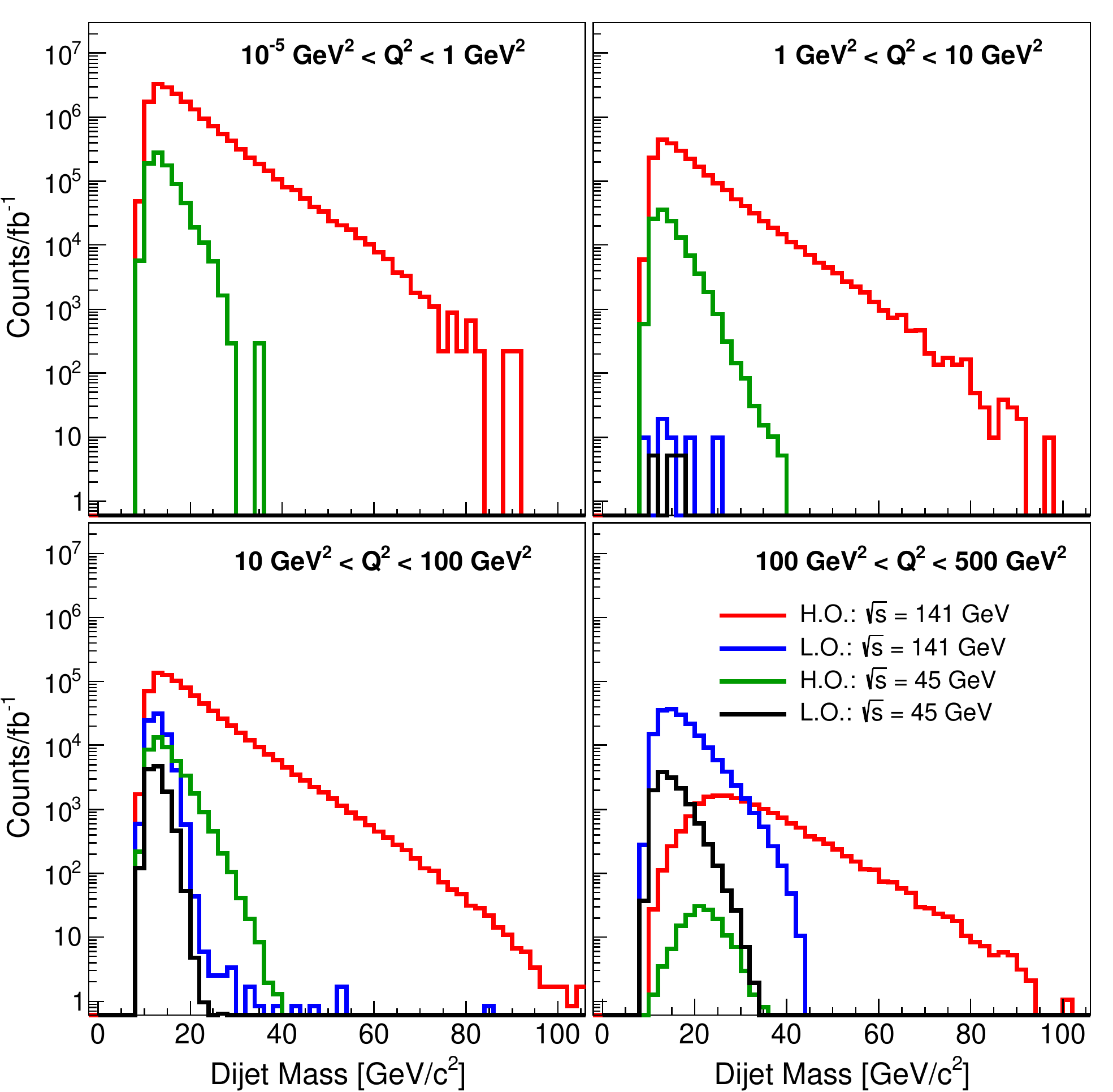}
	\caption[Dijet Mass Spectra]{[color online] Breit frame dijet invariant mass spectra for $\sqrt{s} = 141$ and 45~GeV in \QTWO bins of $10^{-5}-1.0$~GeV$^2$ (upper left), 1-10~GeV$^2$ (upper right), 10-100~GeV$^2$ (bottom left), and 100-500~GeV$^2$ (bottom right). The Resolved, QCDC, and PGF subprocesses have been combined and are compared to the leading order DIS spectra and the histograms have been scaled to the counts expected for an integrated luminosity of 1~fb$^{-1}$.} 
	\label{fig:DIJETMASS}
\end{figure}

The scale relevant for a dijet is its invariant mass, which is simply $\sqrt{(\mathcal{P}_1 + \mathcal{P}_2)^2}$ 
where $\mathcal{P}_1$ and $\mathcal{P}_2$ are the four-momenta of the two jets. The dijet invariant 
mass spectra are shown in Fig. \ref{fig:DIJETMASS} for four \QTWO ranges and center-of-mass energies of 45 and 
141~GeV. The H.O. subprocesses are combined and compared to the L.O. spectra. While the leading order DIS subprocess 
results in only one outgoing quark, at high 
\QTWO, the proton remnant can receive enough transverse momentum to produce a second jet which 
will satisfy the dijet conditions. These L.O. dijets can be effectively separated from the H.O. dijets via a cut on 
the ratio of dijet mass over $Q$ (see Fig. \ref{fig:DISCUT}). 
As was the case with inclusive jet $p_{\mathrm{T}}$, the dijet 
cross section is significantly larger for $\sqrt{s}$ of 141~GeV than for 45~GeV and the spectra extend to 
much higher mass values.

\begin{figure*}[t!]
	\centering \includegraphics[keepaspectratio=true, width=0.75\linewidth]{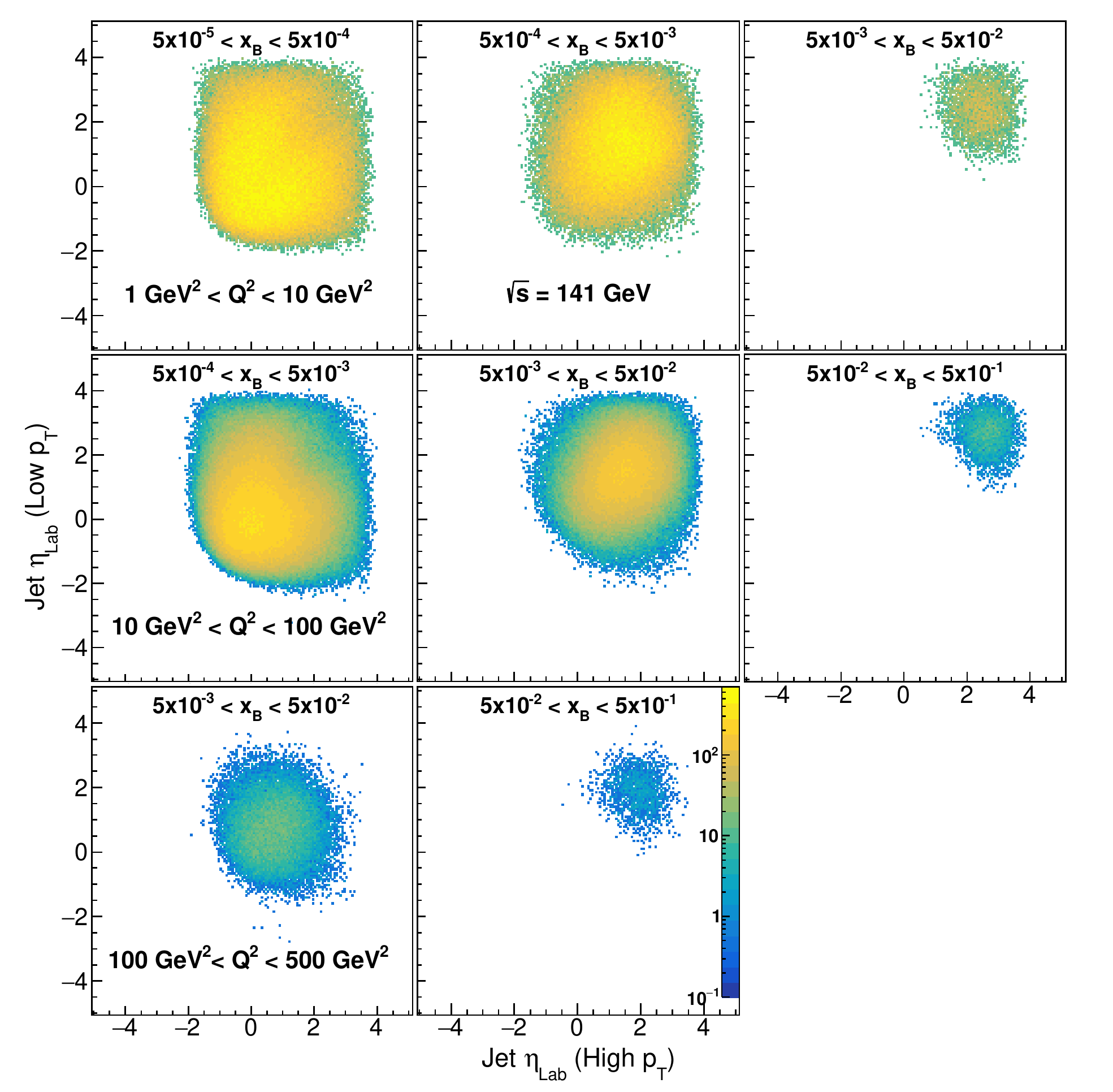}
	\caption[Dijet Eta]{[color online] Laboratory frame jet $\eta - \eta$ correlations for dijets in select \XB bins for \QTWO ranges of 1-10~GeV$^{2}$ (top row), 10-100~GeV$^{2}$ (middle row), and 100-500~GeV$^{2}$ (bottom row). Only the 141~GeV center-of-mass energy is shown. The resolved, QCDC, and PGF subprocesses have been combined and leading order DIS events are not included. Note that all panels share the same scale, which has been set to represent the number of counts expected for an integrated luminosity of 1~fb$^{-1}$.}
	\label{fig:DIJETETAHI}
\end{figure*}

To characterize the location of a dijet in the detector, the pseudorapidities of both jets need to be 
recorded simultaneously as in Fig. \ref{fig:DIJETETAHI}. As before, the H.O. subprocesses have been 
combined and now dijets arising from the L.O. subprocess are not shown. Only $\sqrt{s} = 141$~GeV 
events are shown as the $\sqrt{s} = 45$~GeV distributions are just shifted to lower pseudorapidity for a given 
\XB - \QTWO bin due to the smaller boost from the less energetic hadron beam. As was the case for inclusive jets, jet 
pseudorapidities increase as \XB is increased at a fixed \QTWO, and for a fixed \XB bin, jet pseudorapidities 
decrease as \QTWO is increased.

\begin{figure}[t!]
	\centering \includegraphics[keepaspectratio=true, width=0.95\linewidth]{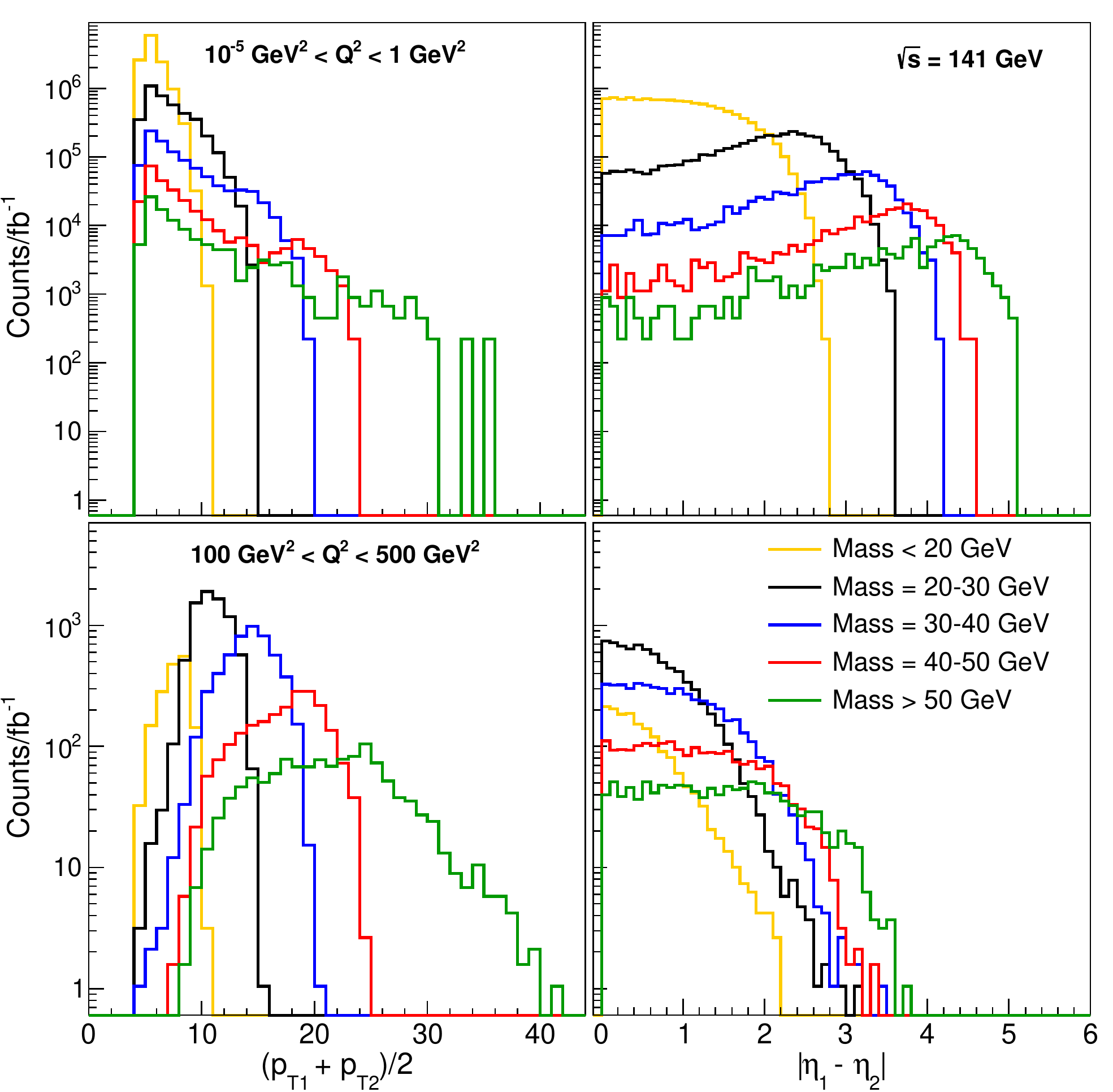}
	\caption[Dijet Average \PT and $\Delta \eta$]{[color online] Average jet \PT (left column) and rapidity difference (right column) for jets comprising a dijet separated in bins of dijet invariant mass for \QTWO ranges of 10-100~GeV$^{2}$ (top row) and 100-500~GeV$^{2}$ (bottom row). Only the 141~GeV center-of-mass energy is shown. The resolved, QCDC, and PGF subprocesses have been combined and leading order DIS events are not included. Histograms have been scaled to the counts expected for an integrated luminosity of 1~fb$^{-1}$.} 
	\label{fig:DIJETPTETA}
\end{figure}

While the absolute pseudorapidities of the two jets comprising a dijet depend on the event \XB and 
\QTWO as shown in Fig. \ref{fig:DIJETETAHI}, the relative pseudorapidity is connected to the dijet invariant mass. 
Expanding the four-vector expression given above, the dijet invariant mass can be approximated 
(ignoring the individual jet masses) as:

\begin{equation} \label{eq:DIJETMASSAPPROX}
	M \approx \sqrt{2p_{T1}p_{T2}\left(\cosh(\Delta \eta) - \cos(\Delta \phi) \right)},
\end{equation}

\noindent where $p_{T1}$ and $p_{T2}$ are the transverse momenta of the two jets and $\Delta \eta$ 
and $\Delta \phi$ are the pseudorapidity and azimuthal angle differences, respectively, between the 
two jets. In this form, it is apparent that dijets can acquire a large mass if their constituent jets have 
large transverse momenta, and/or if the pseudorapidity difference between the two jets is large. The interplay 
between jet \PT and $\Delta \eta$ is made explicit in Fig. \ref{fig:DIJETPTETA} which shows the difference in 
pseudorapidity and average jet 
transverse momenta of the constituent jets for five invariant mass bins and \QTWO ranges of $10^{-5} - 1.0$~GeV$^2$ 
and $100.0 - 500.0$~GeV$^2$. At low $Q^2$, the average jet \PT remains small even for the largest 
mass bins, meaning that larger invariant masses are driven by greater pseudorapidity separations. As \QTWO increases, 
the average jet \PT increases and $\Delta \eta$ does not need to be as large 
to produce a high invariant mass dijet. Thus, even for high mass dijets, it will be important to have good 
jet energy resolution to low \PT and large detector acceptance.

\subsection{Minimum Particle Transverse Momentum} \label{sec:MINPARTPT}

The jets used in the above discussion were created from particles which had a transverse momentum of 
at least 250~MeV/$c$ with respect to the beam. This value was chosen as it is slightly higher than the typical cutoff 
used in \PP jet finding at STAR (see for example \cite{Adam:2019aml}), which when running at $\sqrt{s} = 200$~GeV should have 
relatively similar particle \PT spectra as can be expected at an EIC. The particle \PT cutoff is 
largely driven by detector considerations, with the magnetic field strength often the dominant 
factor. While detector designs for the EIC are still in active development, many include a relatively 
large solenoidal magnetic field of 2 to 3 Tesla in order to provide good \PT resolution over the 
full pseudorapidity range. This will limit the acceptance for low \PT charged particles as they will 
bend so severely in the magnetic field that they will not reach the calorimeters and will be displaced 
significantly from any neutral particles which arise from the hadronizing parton. 

\begin{figure}[t!]
	\centering \includegraphics[keepaspectratio=true, width=0.95\linewidth]{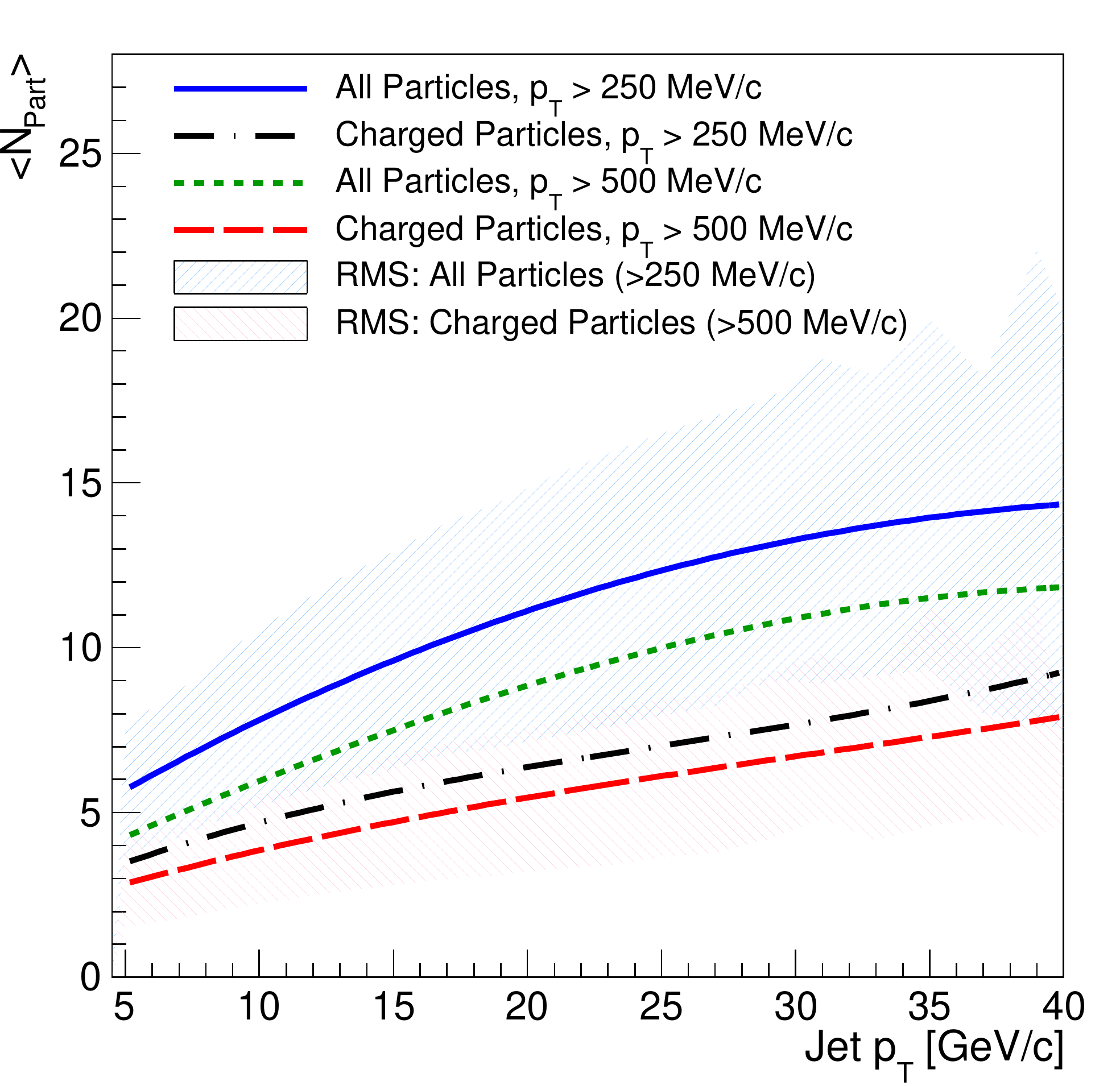}
	\caption[Average Number of Jet Particles Vs Jet \PT]{[color online] The average number of particles in a jet as a function of the transverse momentum of the jet for all stable particles and only charged particles for minimum particle p$_{\mathrm{T}}$s of 250 and 500~MeV/c. Also shown are the RMS variations for all particles with $\PT > 250$~MeV/c and charged particles with $\PT > 500$~MeV/c.} 
	\label{fig:NUMPARTVSPT}
\end{figure}

To study the effect that the loss of low \PT particles will have on jet quantities, the jet finding 
was rerun with the low \PT particle cutoff doubled to 500~MeV/$c$. The most obvious effects of raising 
the cutoff are a reduction in jet/dijet yields and the average number of particles in a jet. The 
jet and dijet yields are reduced by roughly 37\% for $10^{-5} < \QTWO < 1$~GeV$^{2}$ and 20\% for 
$10^{-5} < \QTWO < 500$~GeV$^{2}$. The effect of the minimum \PT cut on jet particle content can be seen in 
Fig. \ref{fig:NUMPARTVSPT} for all particles as well as charged hadrons only.

Removing low \PT particles may also affect how well jets reproduce the kinematics of the underlying 
partons due to the loss of energy contributed by these particles. The impact of this loss 
was studied in the same way as the $R$ dependence in Sec. \ref{sec:JETDEFS}, by comparing the reconstructed 
dijet mass to 
the di-parton invariant mass and by measuring $\Delta R$ between the jet and parton directions. The 
higher minimum particle \PT cut slightly reduces the reconstructed dijet mass versus the true di-parton 
invariant mass, much like what was seen when reducing $R$ in Fig. \ref{fig:RADMASSCOMP}, although the 
magnitude of the effect is not as great. There was no visible change in the $\Delta R$ distributions. As 
detector designs become more advanced, further studies will need to be made to ensure that there is 
sufficient acceptance for low \PT particles. 

\section{Underlying Event Properties} \label{sec:UE}

The underlying event activity, which contributes background energy to jet signals is quantified in this section for jets 
produced from the H.O. subprocesses in the Breit frame.
`Underlying event' (UE) refers to those particles, which do not arise from the outgoing hard-scattered partons and 
can contain contributions from initial and final state radiation (ISR, FSR), beam remnants, and multiple parton interactions (MPI). 
As the QCDC, and PGF subprocesses proceed via a direct $\gamma$ + parton interaction, there is no contribution from MPI and 
the only beam remnant arises from the hadron side. On the other hand, the resolved subprocess is defined by parton + parton 
scattering where one parton is supplied by the photon (see Fig \ref{fig:DISgraph}), and because the photon behaves hadronically, 
it will contribute to the beam remnant and allow for MPI. Because MPI effects are expected to be small for the EIC kinematics and 
because they are difficult to model accurately, the MPI contribution has been disabled in the Monte Carlo.

\begin{figure}[hbt] 
\begin{center}
\includegraphics[width=0.32\textwidth,height=0.18\textheight]{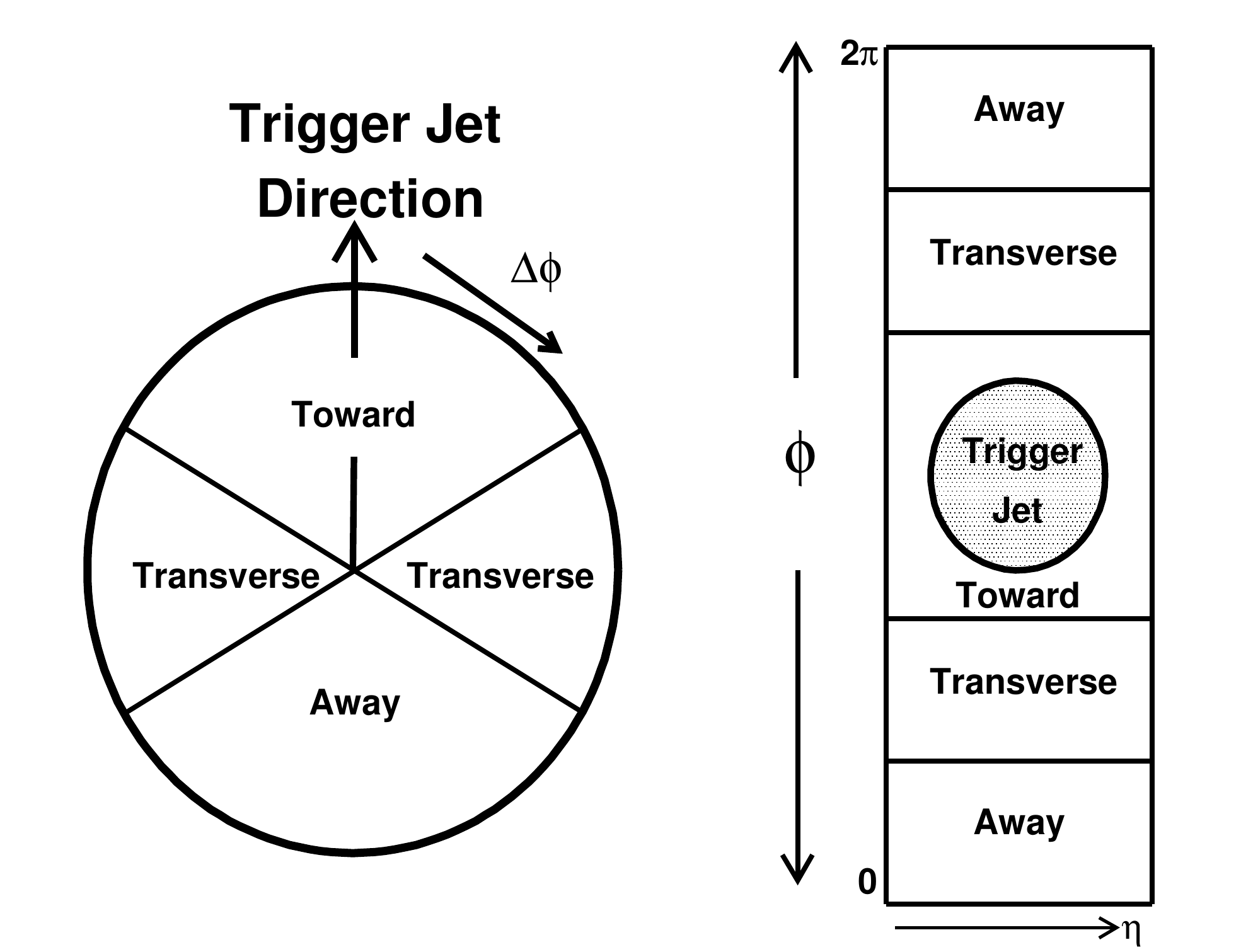} 
\end{center} 
\caption{Illustration of the `Toward', `Away', and `Transverse' regions as defined relative to the highest \PT jet in the event. The angle $\Delta\phi \equiv \phi-\phi_\mathrm{ref~jet}$ is the azimuthal angle between a charged particle and the highest \PT jet, from $-\pi$ to $\pi$. The Toward region is defined as $|\Delta\phi|<60^{\circ}$, while the Away region is $|\Delta\phi|>120^{\circ}$. The Transverse region is defined as $60^{\circ}<|\Delta\phi|<120^{\circ}$. The plot is from~\cite{Affolder:2001xt}.}
\label{fig:UEregion}
\end{figure}

We utilize two methods to analyze UE effects in \EP collisions, the `region method'~\cite{Affolder:2001xt} and 
the `off-axis cone method'~\cite{ALICE:2014dla}. In the region method, the azimuthal angle of the highest \PT jet in each dijet event 
is selected as the reference angle and particles are grouped into one of three regions based on their azimuthal angle 
relative to this reference, $\Delta\phi \equiv (\phi-\phi_\mathrm{ref~jet})$. The particle candidate pool is identical 
to that used in the jet-finding. The `Toward' region is defined as 
$|\Delta\phi|<60^{\circ}$ and contains the reference jet, while the `Away' region has $|\Delta\phi|>120^{\circ}$ and 
generally contains the lower \PT, or associated, jet of the dijet. The `Transverse' region is defined as 
$60^{\circ}<|\Delta\phi|<120^{\circ}$ and the activity here is dominated by the UE. Figure~\ref{fig:UEregion} 
illustrates the definition of the three regions.

\begin{figure*}
\begin{center}
\includegraphics[width=0.68\textwidth]{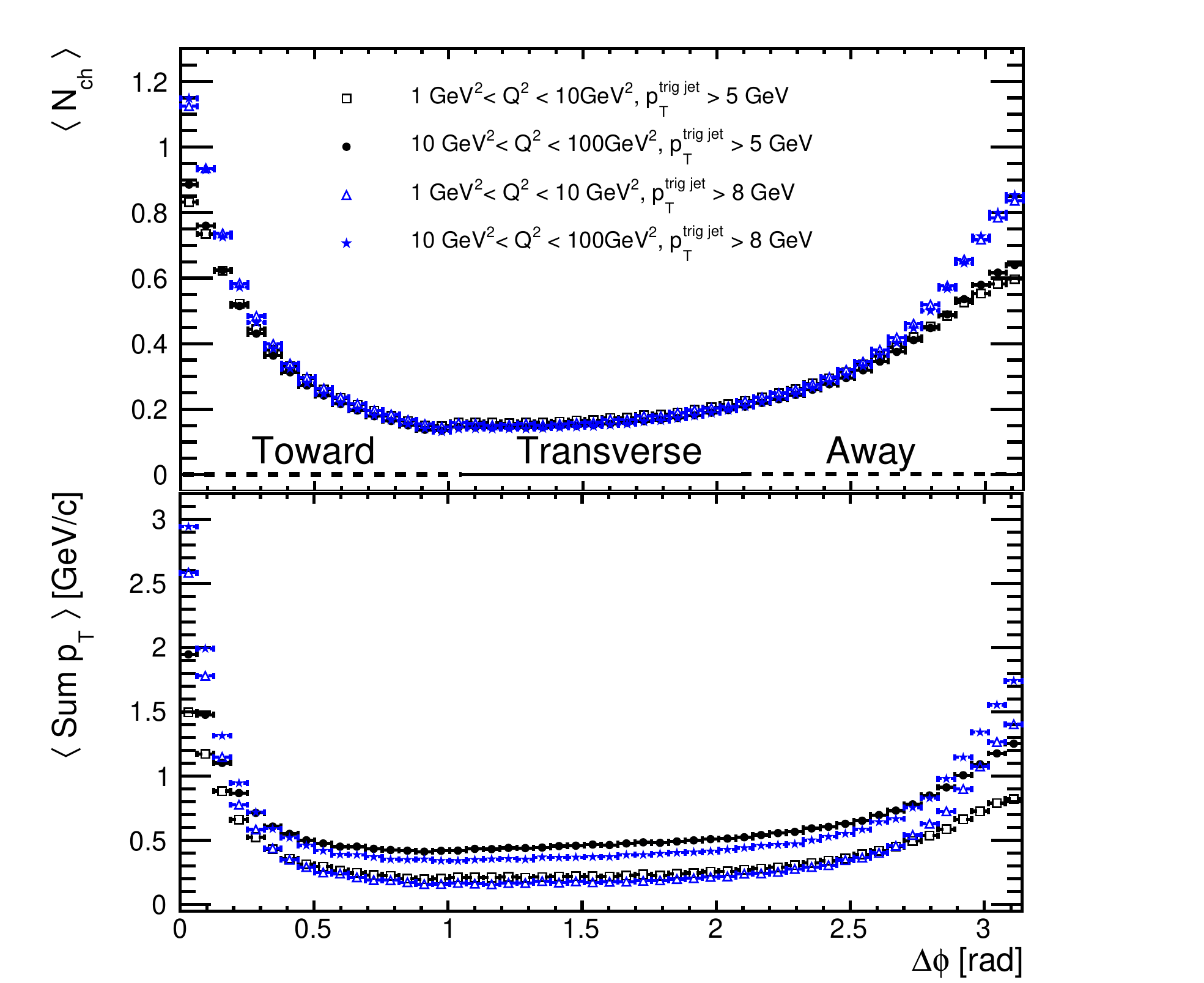}
\end{center} 
\caption{[color online] Average number of charged particles (top) and average charged particle scalar \PT sum (bottom) as a function of the azimuthal angle,
$\Delta\phi$, between the particle and the reference jet for
$p_{T}^\mathrm{ref~jet}>5$ $\mathrm{GeV}$ or $8$ $\mathrm{GeV}$, and
$Q^{2}<10~\mathrm{GeV^{2}}$ or $10<Q^{2}<100~\mathrm{GeV^{2}}$.
Each point corresponds to the $\langle N_{ch} \rangle$ in a
$3.6^{\circ}$ bin.}
\label{fig:UEVsDeltaPhi}
\end{figure*}

Three observables are used to characterize UE activity: the average charged particle multiplicity ($\langle N_{ch} \rangle$), 
the average charged particle scalar \PT sum $(\langle \mathrm{sum}~p_{\mathrm{T}} \rangle)$, and average charged particle \PT $(\langle p_{\mathrm{T}} \rangle)$. Figure \ref{fig:UEVsDeltaPhi} presents 
$\langle N_{ch} \rangle$ (top) and $\langle \mathrm{sum}~\PT \rangle$ (bottom) as a function of $|\Delta\phi|$ for particles 
with $p_{\mathrm{T}}>250$~MeV/c and $-4<\eta<4$. Reference jets with $p_{\mathrm{T}} > 5$~GeV/c (associated jet $p_{\mathrm{T}} > 4$~GeV/c) 
and $p_{\mathrm{T}} > 8$~GeV/c (associated jet $p_{\mathrm{T}} > 7$~GeV/c) were compared along with jets from two \QTWO regions: 
$1~\mathrm{GeV}^2 < Q^2 < 10$~GeV$^2$ and $10~\mathrm{GeV}^2 < Q^2 < 100$~GeV$^2$. There is a strong dependence on jet \PT in the 
Toward and Away regions for both $\langle N_{ch} \rangle$ and $\langle \mathrm{sum}~\PT \rangle$, which is expected as these regions 
are dominated by the jets. In the Transverse region, no dependence on jet \PT is seen for the average number of charged particles 
while a mild difference is seen in the \PT sum for the higher \QTWO range. Interestingly, it is the lower \PT jets which show a 
higher \PT sum in the transverse region, seemingly indicating that jets with higher \PT leave less energy available for underlying 
event activity. No \QTWO dependence is seen for $\langle N_{ch} \rangle$ while larger $Q^2$s lead to greater sum $p_{T}$s in all regions.


\begin{figure*}
\begin{center}
\includegraphics[width=0.68\textwidth]{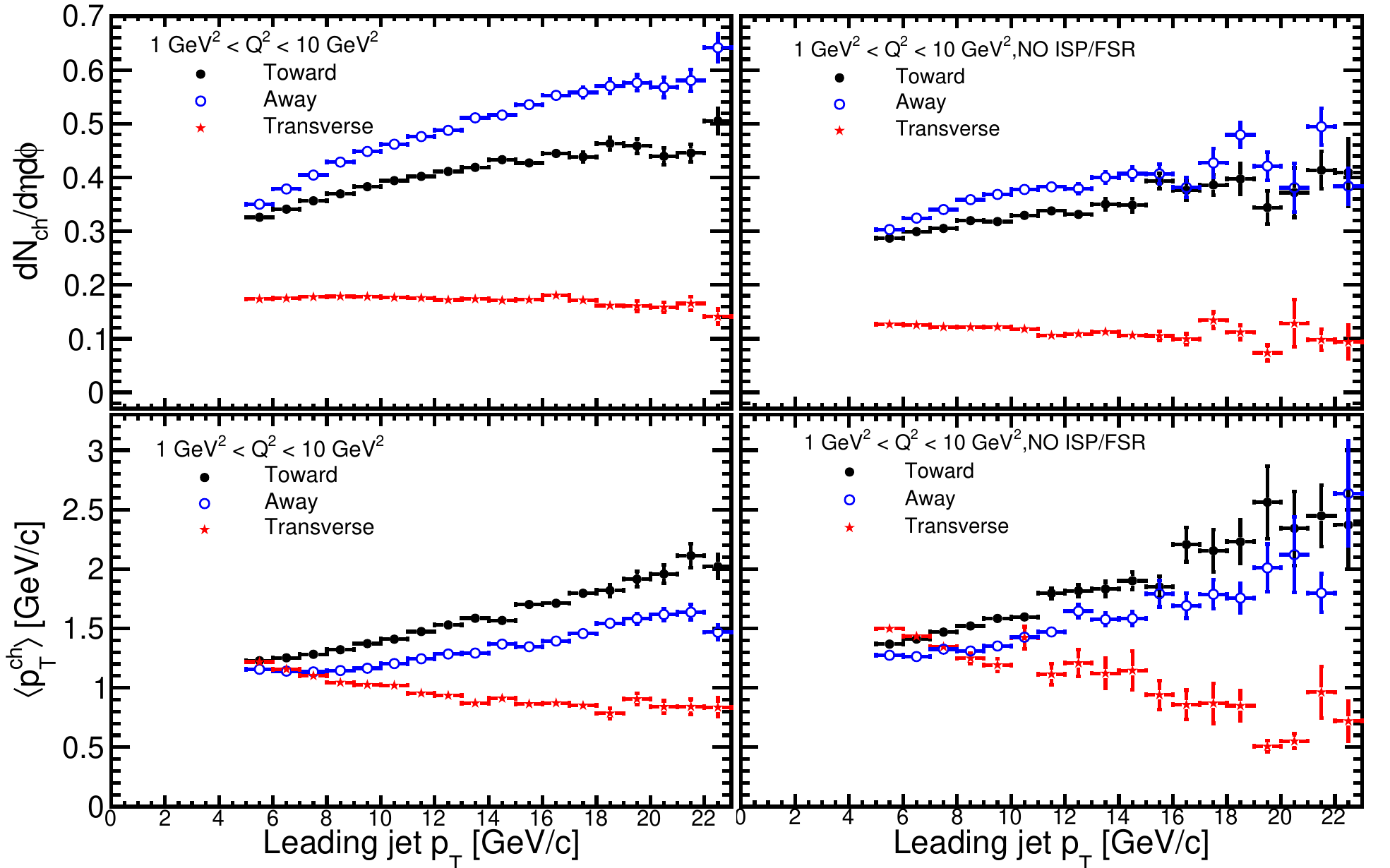}
\end{center} 
\caption{[color online] Charged particle density (top row) and mean \PT (bottom row) as a function of the transverse momentum of the reference jet for the Toward, Away, and Transverse regions. The left column is the standard simulation, while the right column has had initial and final state radiation disabled. Reference jets were required to have $p_{T} > 5$~GeV/c and laboratory pseudorapidity between $\pm 4$ while \QTWO was selected to be between 1~GeV$^2$ and 10~GeV$^2$.}
\label{fig:3Region}
\end{figure*}

The dependence of $\langle N_{ch} \rangle$ and $\langle \PT \rangle$ on the trigger jet \PT in the 
Toward, Away, and Transverse regions is made more explicit in Fig.~\ref{fig:3Region} for \QTWO between 1~GeV$^2$ 
and 10~GeV$^2$. It is seen that both the 
particle density (top) and average \PT (bottom) in the Toward and Away region depend strongly on the trigger jet 
$p_{\mathrm{T}}$ as is to be expected. Conversely, both quantities show a weak anti-correlation with trigger jet 
\PT in the transverse region, in agreement with Fig.~\ref{fig:UEVsDeltaPhi}. The effect of initial and final state 
radiation (ISR/FSR) on the observables 
can be seen in the right hand column of Fig.~\ref{fig:3Region} where the radiation effects have been disabled. The 
presence of ISR/FSR leads to an increase in $\langle N_{ch} \rangle$ for all three regions while surprisingly, the 
average charged particle \PT is seen to increase somewhat without ISR/FSR effects.

\begin{figure*}
	\begin{center}
		\includegraphics[width=0.85\textwidth]{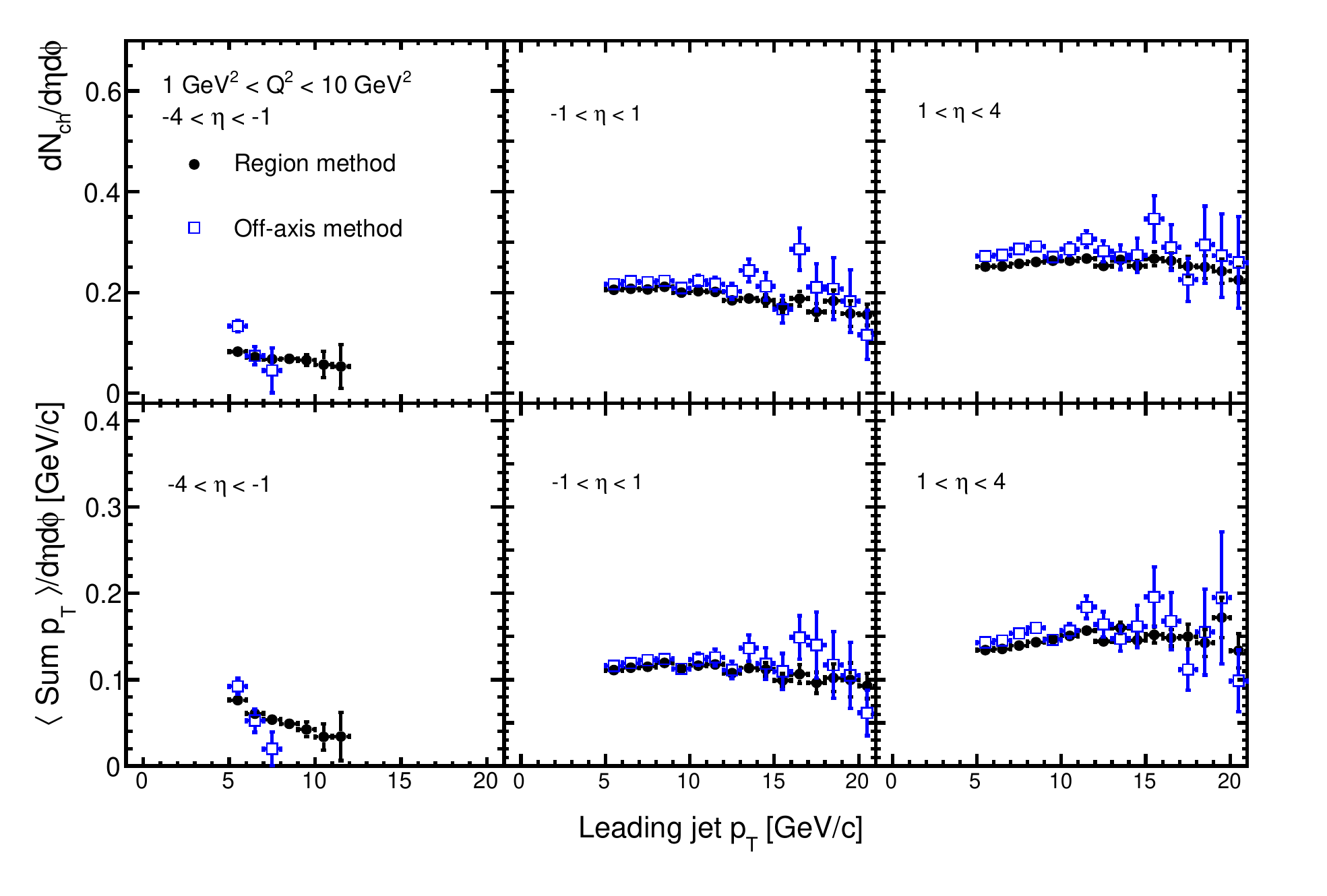} 
	\end{center} 
	\caption{[color online] Average charged particle density (top) and charged particle \PT sum densities (bottom) in the Transverse region as a function of trigger jet \PT for three pseudorapidity ranges and \QTWO between 1~GeV$^2$ and 10~GeV$^2$. Results from the regions 
	method (closed circles) and off-axis cone method (open squares) are compared. The displayed pseudorapidity ranges apply to the particles used in the analysis while the reference jets were required to be more than $0.4$ units of pseudorapidity from a boundary.}
	\label{fig:EtaDep}
\end{figure*}

Unlike the identical species configurations that are often run at colliders, the collisions at an EIC 
will be asymmetric in both particle type and beam energy. This will lead to an asymmetric $\eta$ dependence 
in particle production and UE activity as seen in Fig.~\ref{fig:EtaDep}. Here, the average charged particle 
multiplicity densities and average charged particle \PT sum densities in the Transverse region are shown as a 
function of reference jet \PT for particles in Backward $(-4 < \eta < -1)$, Mid $(-1 < \eta < 1)$, and 
Forward $(1 < \eta < 4)$ pseudorapidity ranges (as defined in the laboratory frame) from the region method 
(filled symbols). The reference jet was required to be within the same pseudorapidity range as the particles with 
the added restriction that the jet $\eta$ must be $0.4$ units away from a range boundary in order to facilitate 
comparisons to the off-axis cone UE characterization method. It is seen that the UE charged particle density is higher 
in the Forward (hadron-going) direction which is expected as any beam remnant contribution will generally 
follow the struck hadron. Also, particle density in the forward region will be higher due to the 
boost from the more energetic hadron beam.

\begin{figure}[hbt] 
	\begin{center}
		\includegraphics[width=0.45\textwidth]{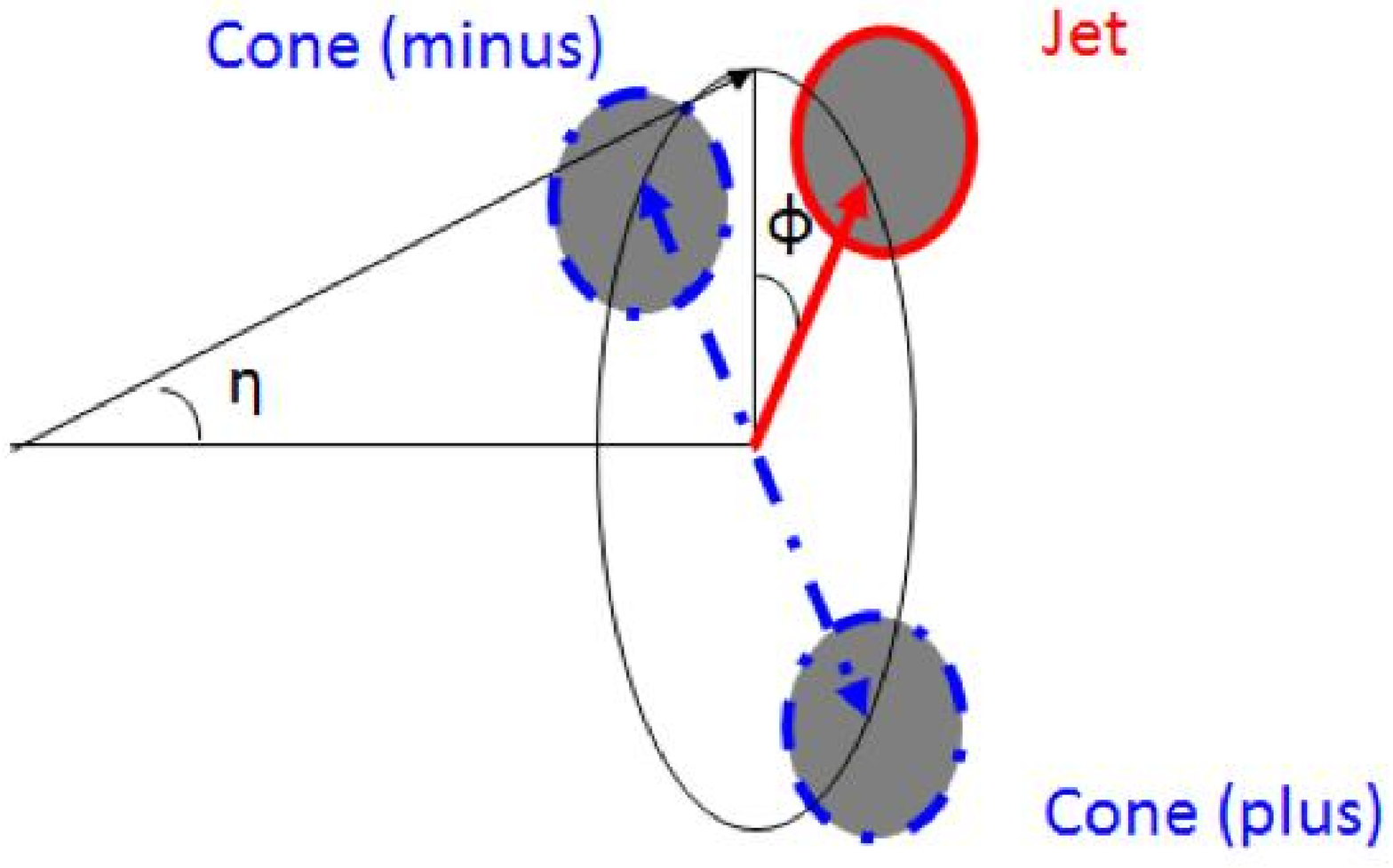} 
	\end{center} 
	\caption{[color online] Illustration of two off-axis cones relative to a jet.}
	\label{fig:UEcone}
\end{figure}

The second technique used to investigate UE effects is the off-axis cone method~\cite{ALICE:2014dla}, developed by 
the ALICE collaboration. The off-axis cone method studies the UE on a jet-by-jet level, as opposed to the region 
method which is designed to study the UE on the event level. For every reconstructed jet, two off-axis cones 
(cone(-) and cone(+)) are defined, each of which is centered at the same $\eta$ as the jet but $\pm\pi/2$ away in 
$\phi$ from the jet $\phi$, as shown in Fig.~\ref{fig:UEcone}, and the particles which fall inside these cones are 
used to characterize the underlying event. The cone radius was chosen to be $0.4$ so as not to overlap with the 
primary jet. The multiplicity density is defined as the average number of charged particles inside each cone, 
$\langle N_{ch} \rangle$, divided by the cone area while the $\langle p_{T}~\mathrm{sum} \rangle$ 
density is defined as the average off-axis cone $p_{T}$ divided by the cone area. 

The UE results from the off-axis method (open symbols) are compared to the region method (closed symbols) in 
Fig.~\ref{fig:EtaDep}. As with the reference jets from the region method, the jets from the off-axis method were  required to be more than $0.4$ away from a range boundary in $\eta$ so that the full off-axis cone would fit in 
the indicated pseudorapidity bin. With the jet and particle pseudorapidities defined in this way, a faithful 
comparison between the region and off-axis methods can be made and the agreement is seen to be very good. The 
pseudorapidity dependence of the UE activity seen in Fig.~\ref{fig:EtaDep} means the off-axis cone method will be 
important when correcting jet quantities for underlying event contamination as it will be necessary to measure that 
component at the pseudorapidity of the jet.

The results described above characterize the expected underlying event at an EIC as generated in our Monte Carlo. 
To get a better feeling for the size of these effects and provide a sanity check on the simulation, it is 
instructive to compare the simulated \EP results with \PP data at a similar center-of-mass energy. The STAR 
experiment \cite{Ackermann:2002ad} at the Relativistic Heavy Ion Collider (RHIC) has 
performed a similar analysis of UE activity as presented here on \PP data taken at $\sqrt{s} = 200$~GeV, which 
provides an opportunity for such a comparison. The STAR analysis~\cite{Adam:2019xpp} measured both 
$\langle N_{ch} \rangle$ and $\langle \PT \rangle$ for all charged particles in the mid-rapidity region 
$(-1 < \eta < 1)$ with $p_{\mathrm{T}} > 0.2$~GeV/c. The STAR average charged particle density varies from 0.8 
to 0.5 for reference jets with \PT of 5~GeV/c to 40~GeV/c. This is a factor of roughly two to four greater than 
what is observed at mid or forward rapidity in Fig.~\ref{fig:EtaDep}. This is not surprising as the STAR result 
involves the collision of two protons. However, the average charged particle \PT measured by STAR is relatively flat 
with a value of 0.6~GeV/c, which is at least a factor of two lower than the result presented in 
Fig.~\ref{fig:3Region}. The larger $\langle \PT \rangle$ in the Transverse region at the EIC is due to the the 
boost into the Breit frame. While the particles which participate in the hard scattering processes initially move along the 
photon-parton axis, the underlying event particles arise largely from the proton and are thus more aligned with the beam 
axis. When measured with respect to the photon-parton axis, as is done in the Breit frame, these underlying event particles 
acquire, on average, larger transverse momenta. When analyzed in the laboratory frame, $\langle \PT \rangle$ in the Transverse region 
is roughly 0.6~GeV/c, in agreement with the STAR result.

\section{Detector Effects} \label{sec:DETECTOR}

The jets used for the results presented in Secs. \ref{sec:JETFINDING} and \ref{sec:UE} were reconstructed at 
`particle level', taking as input the exact four-momenta of all generated final state particles. 
These jets do not include distortions which will arise from the finite energy and momentum 
resolutions and inefficiencies of any real detector. 
Because the entirety of the EIC physics program requires high resolution calorimetry and tracking 
over a wide acceptance range, the induced distortions are expected to be small. Nevertheless, 
it is important 
to investigate how jets will be affected by a realistic detector environment. 
In order to quantify how these distortions will affect jet reconstruction, the energy and 
momenta of input particles were smeared based on a model EIC detector before being 
clustered into jets. These smeared jets were then compared to the corresponding particle level jets 
to study detector effects. 

\subsection{Smearing Generator and Detector Model}

Generally, detector effects are investigated by propagating simulated events through a detailed 
detector model which reproduces the relevant energy and momentum resolutions, efficiencies, material 
budgets, and readout responses of the actual device. As such detailed models for prospective EIC 
detectors are only starting to be developed, and key detector technology choices are still in flux, 
the effects of finite resolution and acceptance on jet-finding were explored using a smearing 
generator, which alters a particle's energy or momentum based on a specific resolution function. 
While not a substitute for a full detector simulation, this smearing method has the benefit of 
being much faster computationally, making it easy to investigate different sub-detector 
configurations and resolutions.

The smearing generator used allows a user to define `devices' which encode the behavior of 
individual or collections of detector subsystems. A single device will smear the energy, momentum, 
or direction of all particles which fall into its acceptance. Here, acceptance not only 
refers to the spatial extent of the device, but also to particle properties such as charge and 
how the particle interacts with the detector material (hadronically or electromagnetically). 
Three particle charge and interaction types are used in the smearing performed here: charged hadronic, 
neutral hadronic, and electromagnetic.  Charged hadronic particles (assumed to be detected using a tracker) have 
their momenta and trajectories smeared while neutral hadronic and electromagnetic 
particles (assumed to be detected with calorimeters) have their energies smeared. Because a device will only 
smear either the energy or momentum component of a particle, the energy-momentum-mass relationship 
of the smeared 4-vector will be broken. To address this, after the smearing was performed, the 
charged hadron energies were altered to match their momenta assuming the particles had a pion mass. 
Similarly, the momenta of neutral hadrons and particles interacting electromagnetically were set equal 
to the smeared energy, which is equivalent to assuming the particle was massless. This simplistic 
compensation will be inadequate for particles with significant mass, such as protons and neutrons, but is 
sufficient for the purpose of this study. When more complete detector simulations are developed, efforts 
should be made to determine the utility of the calorimeter systems as well as particle identification 
for more accurate particle four-momentum reconstruction. 

\begin{figure}[t!]
 \centering \includegraphics[keepaspectratio=true, width=0.95\linewidth]{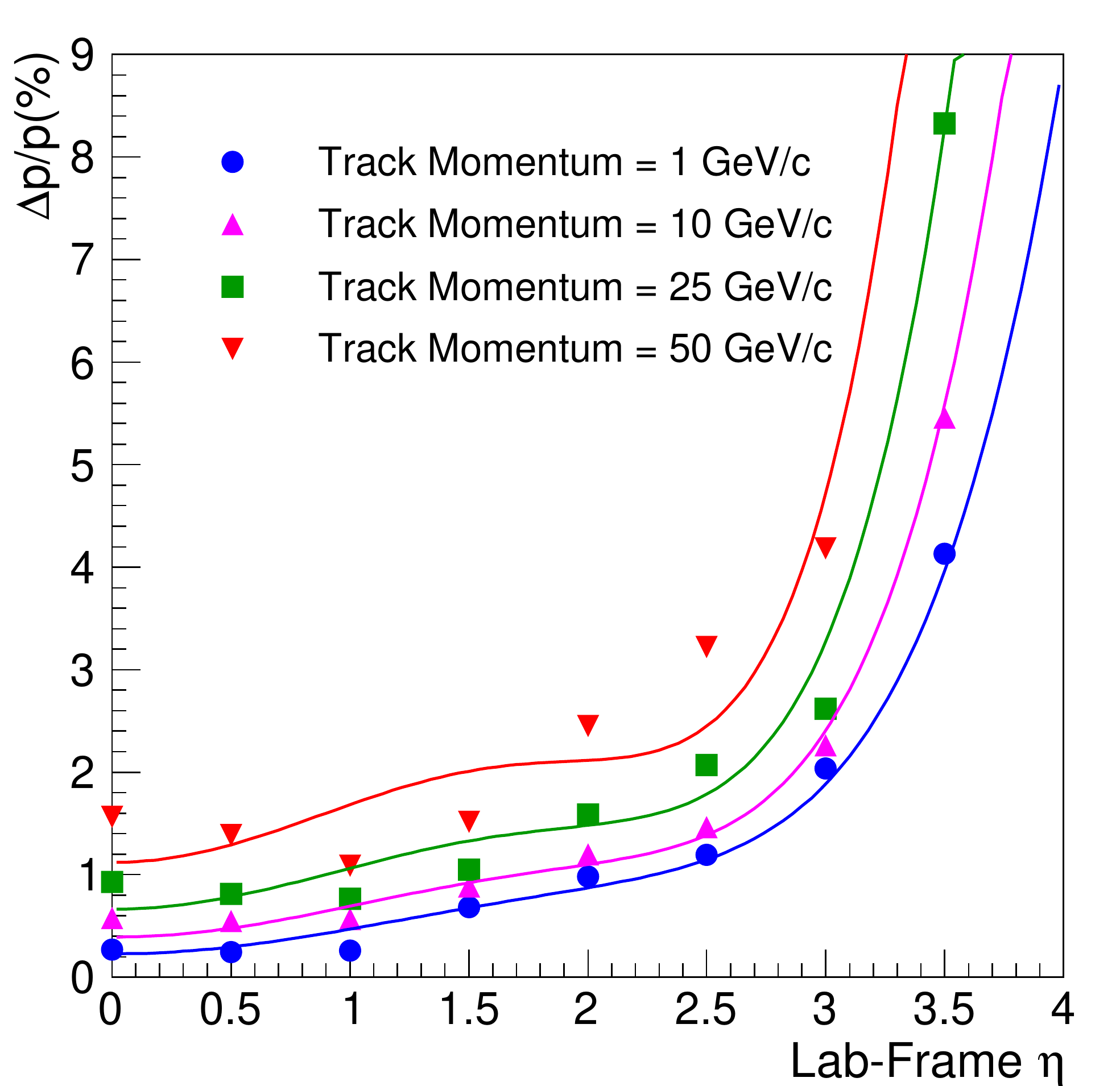}
 \caption[Track Momentum Resolution]{[color online] Track momentum resolution assumed for the smearing generator as a function of track pseudorapidity. The points represent extractions of resolution at specific momenta and pseudorapidity from simulation of a model BeAST tracking detector and the curves are instances of the function used to fit the points that was passed to the smearing generator.} 
	\label{fig:TRACKMOMRES}
\end{figure}

For this study, the smearing generator devices were defined such that they would reproduce 
the projected behavior of BeAST, Brookhaven's `green field' detector proposal. 
BeAST is built around a 3 Tesla solenoidal magnet and will include high precision tracking 
detectors spanning a pseudorapidity range of $|\eta| < 3.5$, electromagnetic calorimetry covering 
the range $|\eta| < 4.0$, and hadron calorimetry in the forward and backward regions 
$1 < |\eta| < 4.0$. BeAST will also have good vertex detection and particle identification 
capabilities as well as instrumentation to detect particles scattered at small angles, both in the 
hadron and lepton beam direction, such as Roman pots and a system to tag low $Q^2$ electrons.
However, these systems do not directly affect jet reconstruction and were therefore not included 
in this simulation. The calorimeter resolutions assumed for different detector regions can be found 
in Tab. \ref{tab:CALRES} while the tracking resolution for different particle momenta as a function 
of pseudorapidity can be seen in Fig. \ref{fig:TRACKMOMRES}. 
Several modifications to the baseline BeAST configuration were also considered, including the 
introduction of a track finding inefficiency factor of 5\% and the addition of a mid-rapidity 
hadron calorimeter assuming a high and low energy resolution.

\begin{table}[ht!]
	\setlength{\tabcolsep}{9.0pt}
	\begin{center}
		\scalebox{0.9}{
			\begin{tabular}{ c  c  c }
				\hline \hline
				Component & Pseudorapidity Range & Resolution \\ \hline
				Back EMCal & $-4.0 < \eta < -2$ & $\frac{1.5\%}{\sqrt{E}} \oplus 1\%$ \\ 
				Mid-Back EMCal & $-2 < \eta < -1$ & $\frac{7\%}{\sqrt{E}} \oplus 1\%$ \\ 
				Mid EMCal & $-1 < \eta < 1$ & $\frac{10\%}{\sqrt{E}} \oplus 1\%$ \\ 
				Fwd EMCal & $1 < \eta < 4.0$ & $\frac{10\%}{\sqrt{E}} \oplus 1\%$ \\ 
				Fwd/Back HCal & $1 < |\eta| < 4.0$ & $\frac{50\%}{\sqrt{E}} \oplus 10.0\%$ \\ 
				Lo Res Mid Hcal & $-1 < \eta < 1$ & $\frac{75\%}{\sqrt{E}} \oplus 15\%$ \\ 
				Hi Res Mid Hcal & $-1 < \eta < 1$ & $\frac{35\%}{\sqrt{E}} \oplus 2\%$ \\ \hline \hline
			\end{tabular}
		}
		\caption[Calorimeter Resolutions]{Assumed energy resolutions and psuedorapidity ranges for the electromagnetic and hadron calorimeters included in the detector smearing model.}
		\label{tab:CALRES}
	\end{center}
\end{table}

\subsection{Smearing Results}

Using the smearing procedure and resolution parameters presented above, individual particle 
4-momenta were altered and then passed to the jet-finder to be clustered into jets using the 
same procedure as for unaltered particles. 
Thus, for each event, there will be a set of unaltered particle level jets and a set 
of smeared jets. In order to evaluate how the smearing procedure has modified the properties of a 
given particle level jet, an association must be made between that particle level jet and a 
particular smeared jet. This is done by finding the smeared jet which minimizes the quantity 
$\Delta R = \sqrt{(y_{\mathrm{Particle}} - y_{\mathrm{Smeared}})^2 + (\phi_{\mathrm{Particle}} - \phi_{\mathrm{Smeared}})^2}$ 
for each particle level jet, with $y$ being the rapidity and $\phi$ the azimuthal angle of the jet. Particle level 
and smeared jets were required to have $\Delta R < 1.0$ in order to be considered associated.

\begin{figure}[t!]
 \centering \includegraphics[keepaspectratio=true, width=0.95\linewidth]{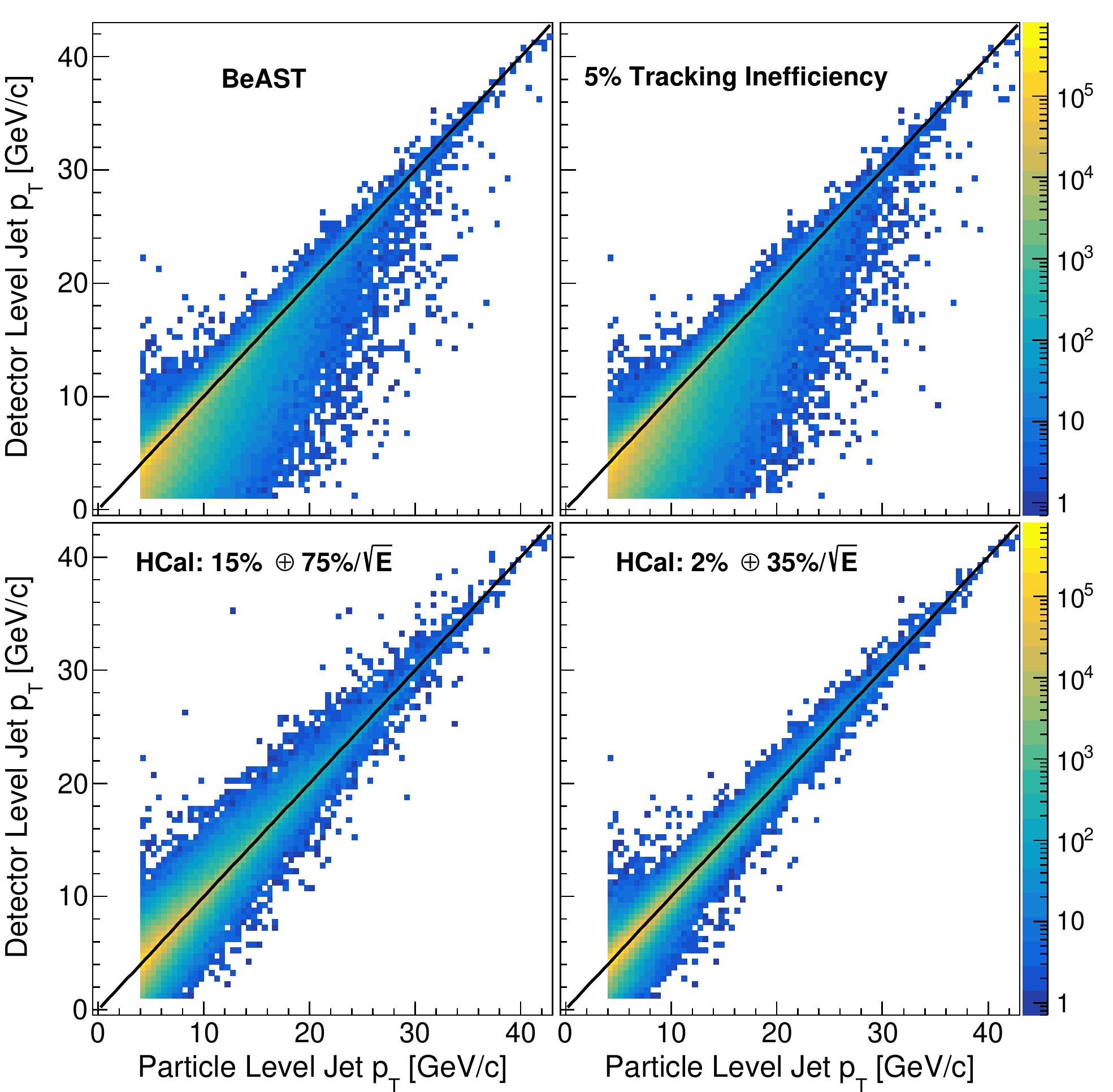}
 \caption[Jet Transverse Momentum Correlation]{[color online] Correlation between particle level and smeared 
          jet $p_{T}$ for the BeAST detector setup (upper left), BeAST assuming a 5\% track finding inefficiency (upper right), 
          and BeAST assuming a mid-rapidity hadron calorimeter with resolution $15\% \oplus \frac{75\%}{\sqrt{E}}$ (lower left) and resolution $2\% \oplus \frac{35\%}{\sqrt{E}}$ (lower right).} 
	\label{fig:JETDETVSPARPT}
\end{figure}

The relationship between the transverse momenta of associated particle level and smeared jets can 
be seen in Fig.  \ref{fig:JETDETVSPARPT} for the baseline BeAST design, as well as the 5\% track 
finding inefficiency and mid-rapidity hadron calorimeter scenarios. As the baseline design does 
not include a hadron calorimeter at mid-rapidity, neutrons and $K^{0}_{\mathrm{L}}$'s in this 
region are not detected, meaning smeared jets will tend to have lower transverse momentum than 
their corresponding particle level jets. The population and extent of this tail depends on the 
average number of neutral hadrons in the event sample and the amount of transverse momentum they carry. 
Removing 5\% of charged particles increases somewhat the number of events which populate the 
off-diagonal tail. The inclusion of a hadron calorimeter at mid-rapidity captures the 
remaining neutral energy making the correlation between particle level and smeared jet $p_{T}$ 
more symmetric around the diagonal. 
The width of the distribution is then determined by the resolution of the hadron calorimeter.

\begin{figure*}[t!]
 \centering \includegraphics[keepaspectratio=true, width=0.95\linewidth]{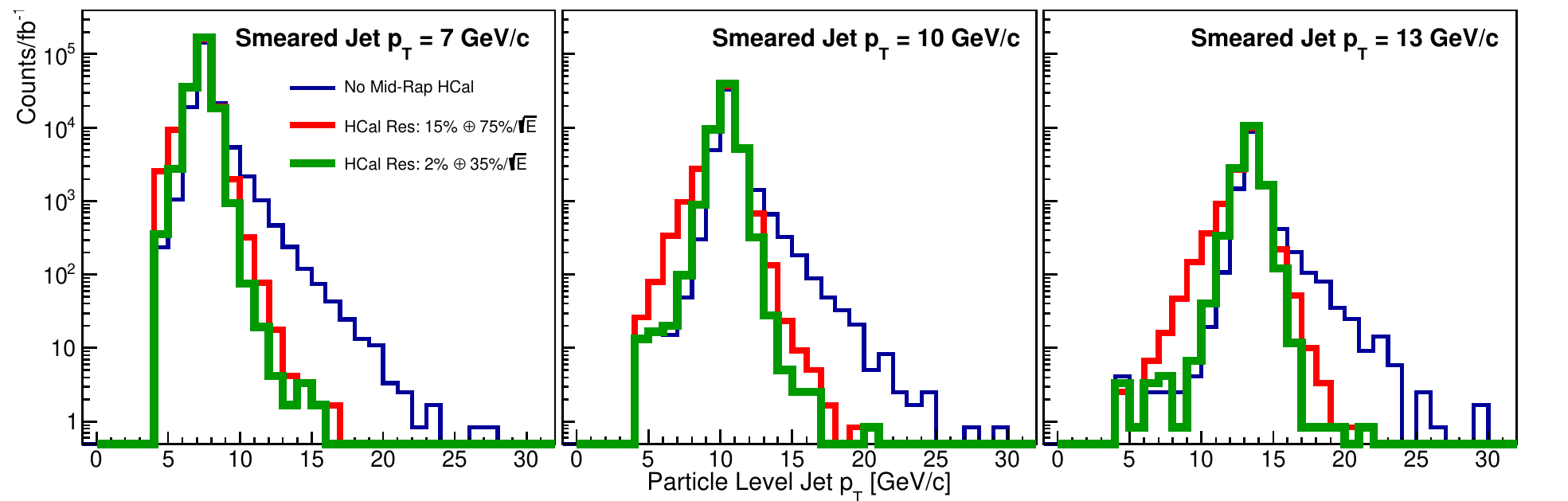}
 \caption[Detector Vs Smeared Projections]{[color online] Particle jet $p_{T}$ spectra for smeared jet $p_{T}$ 
          values of 7 (left), 10 (middle), and 13~GeV/c (right). The baseline BeAST design (blue) and BeAST plus two 
          mid-rapidity hadron calorimeter configurations (red and green) are compared.}
	\label{fig:PROJECTIONS}
\end{figure*}

A more detailed comparison of the relationships shown in Fig. \ref{fig:JETDETVSPARPT} can be 
obtained by taking projections onto the particle level axis for narrow slices of smeared jet 
\PT (or vice versa). Figure \ref{fig:PROJECTIONS} presents three such projections with 
smeared jet transverse momenta of 7, 10, and 13~GeV/$c$ for the baseline BeAST and BeAST with two
mid-rapidity hadron calorimeter configurations and essentially shows how different particle level 
\PT values contribute to a given smeared jet \PT. It is evident that the high resolution hadron 
calorimeter (green) substantially improves the jet resolution, however, it is less clear that the 
low resolution calorimeter (red) provides much advantage over the baseline design (blue). 
Smeared jets found with the low resolution calorimeter option have less contribution from  
particle level jets with larger \PT than in the baseline design, however, these smeared jets obtain a 
large contribution from lower \PT particle level jets due to the large energy distortion 
introduced by the calorimeter. Implications of this observation will be discussed in the next section.

In addition to transverse momentum, smearing of the rapidity and azimuthal angle of the jet thrust 
axes were also investigated and very good agreement between particle level and smeared jets was seen. 
It should be noted, however, that the position resolutions inherent to the calorimeters were not 
considered in this exercise as they depend on details such as material, tower size, and readout which 
have not been finalized. This should be revisited when more complete detector simulations are available 
and will be critically important for future work investigating the utility of jet shape observables.

\subsection{Hadron Calorimetry} \label{sec:HCAL}

The decision not to include a mid-rapidity hadron calorimeter in the BeAST design was based on 
several considerations including the low energies of produced particles, the modest fraction 
of total energy carried by neutral hadrons, the use of streaming readouts which do not require 
a trigger, and finally, the significant cost of such a detector. 
Figure \ref{fig:PROJECTIONS} makes it clear that a hadron calorimeter with sufficiently high 
resolution can markedly improve jet energy measurements. 
Unfortunately, calorimeter cost 
increases with resolution, meaning the inclusion of such a high resolution mid-rapidity hadron 
calorimeter (which must cover a large volume) may be infeasible. While the corrections to jet 
energy needed in the absence of a hadron calorimeter will be modest, it is worth considering 
the benefits that could be provided by a more economical lower resolution calorimeter.

One such benefit would be the ability to implement an unbiased jet trigger. While the current 
plan calls for a data acquisition system capable of recording all interactions, this ability 
has not been demonstrated, which means that the capability of triggering on events with jets 
in an unbiasd way could be necessary. Even if such a streaming readout is possible, a traditional 
trigger system including a hadron calorimeter may be more economically feasible. 
A hadron calorimeter would also provide \textit{in situ} 
measurements of neutral hadron abundances and energies, which would reduce the uncertainty 
in any Monte Carlo based corrections to the jet energy. Figure \ref{fig:JETDETVSPARPT} shows that 
even a low resolution hadron calorimeter would reduce the number of jets reconstructed at significantly 
lower transverse momenta, which would mitigate the loss of jets which would otherwise fail a minimum 
\PT cut. The largest benefit, though, would 
likely come from the ability to differentiate between jets which do and do not contain neutral 
hadrons.

\begin{figure}[t!]
 \centering \includegraphics[keepaspectratio=true, width=0.75\linewidth]{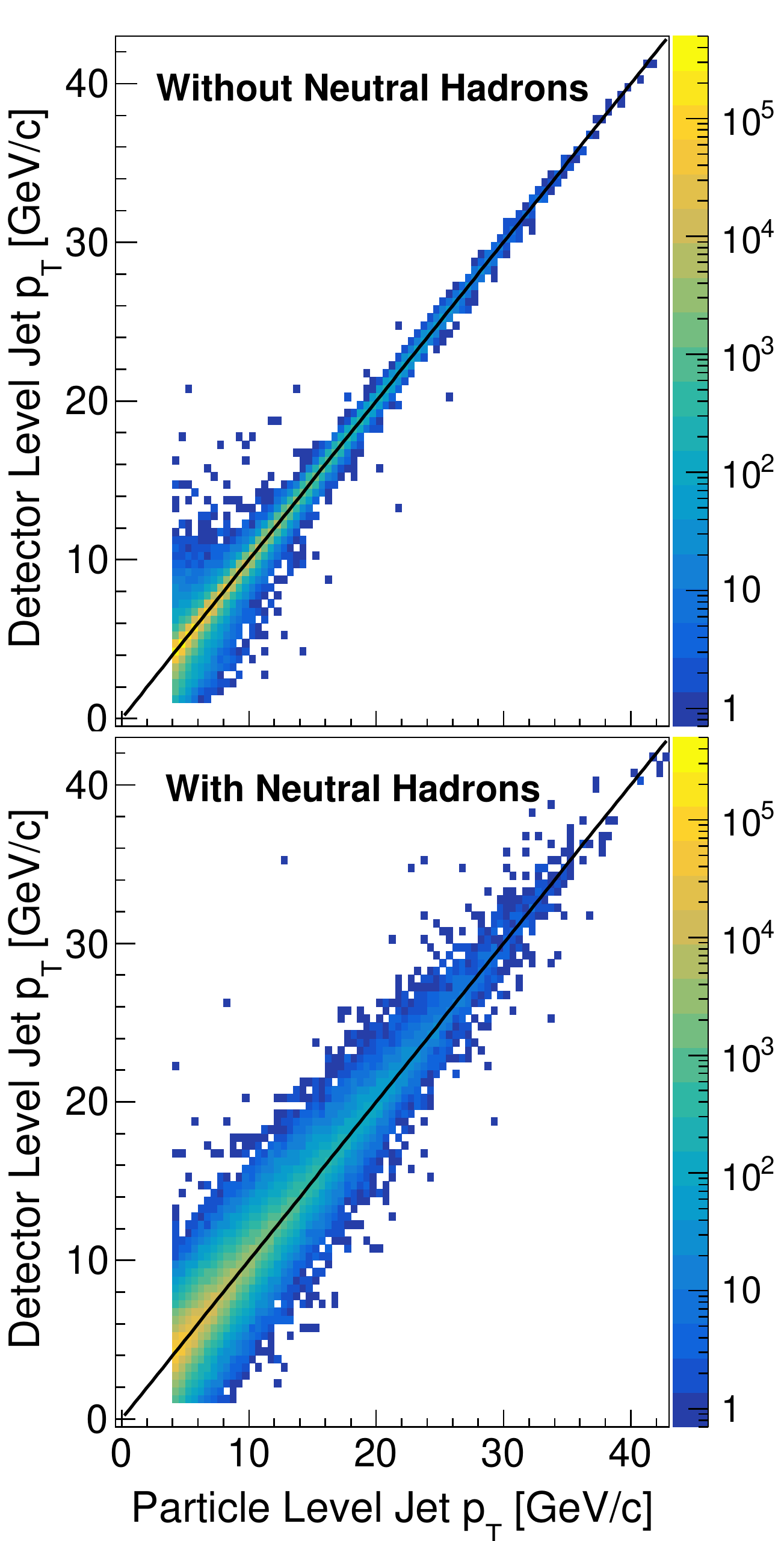}
 \caption[With and Without Neutral Hadrons]{[color online] Relationship between particle level and smeared jet transverse momenta for jets which do not (upper panel) and do (lower panel) contain neutral hadrons.}
	\label{fig:NEUTRALCOMP}
\end{figure}

A hadron calorimeter should make it possible to separate jets containing neutral hadrons from 
those which don't by identifying energy deposits which do not have a corresponding charged 
particle track. The energy resolutions of the roughly 65\% of jets which do not contain a 
neutral hadron will be dominated by the high precision tracker and electromagnetic calorimeters. 
The superior resolution for jets which do not contain neutral hadrons versus those which do, 
can be seen in Fig. \ref{fig:NEUTRALCOMP}. Separating jets in this way would allow a much smaller 
correction to be applied to the majority of jets while reserving the larger corrections for the 
35\% of jets which contain energy from neutral hadrons. This scheme should improve overall jet 
energy resolution much more than what would be possible considering only the energy recorded by 
the calorimeter.

\section{Jet Application: Tagging Photon-Gluon Fusion} \label{sec:DELTAG}

Previous sections have focused on technical aspects of jet finding at an EIC without discussion 
of potential applications. 
As stated above, the utility of dijets at an EIC has been explored recently in the context of accessing the gluon 
Sivers function \cite{Zheng:2018awe} and Weiz\"{a}cker-Williams gluon distributions \cite{Dumitru:2018kuw} as well as determining polarized and unpolarized photon structure functions \cite{Chu:2017mnm}. 
This section will present a related measurement in which dijets are used to tag photon-gluon 
fusion events for the purpose of exploring the gluon contribution to the spin of the proton, 
\DG, via the longitudinal double spin asymmetry \ALL at leading order. 

\subsection{Kinematics and Tagging}

One of the signatures of the PGF process is the production of particles with large momenta 
transverse to the photon-proton interaction axis which are back-to-back in azimuth, meaning the 
observation of a dijet in the Breit frame can be used to tag possible PGF events. Unfortunately, 
both the resolved and QCD compton processes, which are background to a \DG measurement, 
also produce such dijets. 
While a global analysis could likely handle these background contributions in a consistent way, 
it is worth exploring what can be done experimentally to isolate the PGF process. 

Because the dijet kinematics approximate those of the outgoing partons, they can be used 
to reconstruct properties of the event which will help separate the PGF process from resolved 
and QCDC events. In this analysis, the two variables used for this purpose are $x_{\gamma}$ 
and $x_{p}$, which are the momentum fractions carried by the parton originating from the photon 
and the parton coming from the proton, respectively. 
These quantities are reconstructed from the dijet kinematics as follows:

\begin{align} \label{eq:XGAMMAPROTON}
x_{\gamma} &= \frac{1}{2E_{e}y}\left(m_{T1}e^{-Y_{1}} + m_{T2}e^{-Y_{2}} \right) \\
x_{p} &= \frac{1}{2E_{p}}\left(m_{T1}e^{Y_{1}} + m_{T2}e^{Y_{2}} \right),
\end{align}

\noindent where $E_{e}$ and $E_{p}$ are the energies of the incoming electron and proton beams, 
respectively, $y$ is the inelasticity, $m_{T}$ is the jet transverse mass defined as the 
quadrature sum of the jet mass and \PT, and $Y$ is the jet rapidity, in the laboratory frame. 
The correlation between generated and reconstructed $x_{\gamma}$ and $x_{p}$ is quite good 
(see \cite{Chu:2017mnm}).
Dijets were reconstructed using the same method as described in Sec. \ref{sec:DIKIN}. 

\begin{figure*}[t!]
	\centering \includegraphics[keepaspectratio=true, width=0.95\linewidth]{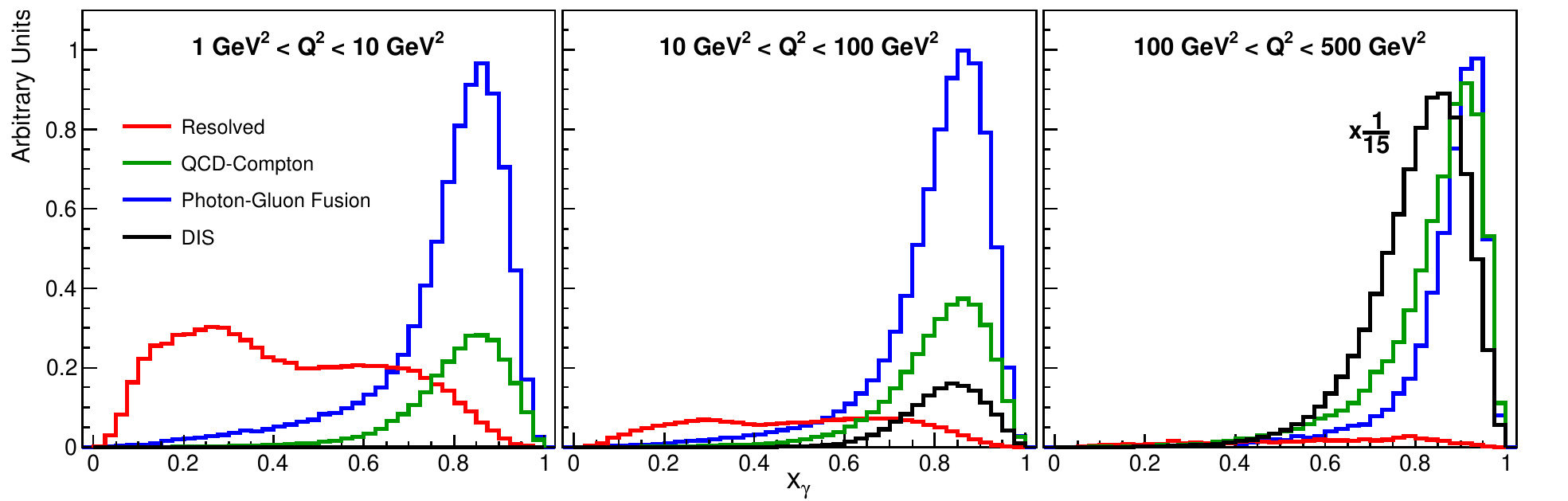}
	\caption[Reconstructed $x_{\gamma}$ by Subprocess]{[color online] Reconstructed $x_\gamma$ for the resolved, QCDC, PGF , and DIS subprocesses in \QTWO bins of 1-10~GeV$^2$ (left), 10-100~GeV$^2$ (middle), and 100-500~GeV$^2$ (right). Note that each panel has been scaled separately an arbitrary amount and in the right panel, the DIS curve has been scaled down by an additional factor of 15.}
	\label{fig:RECOXGAMMA}
\end{figure*}

\begin{figure}[t!]
	\centering \includegraphics[keepaspectratio=true, width=0.75\linewidth]{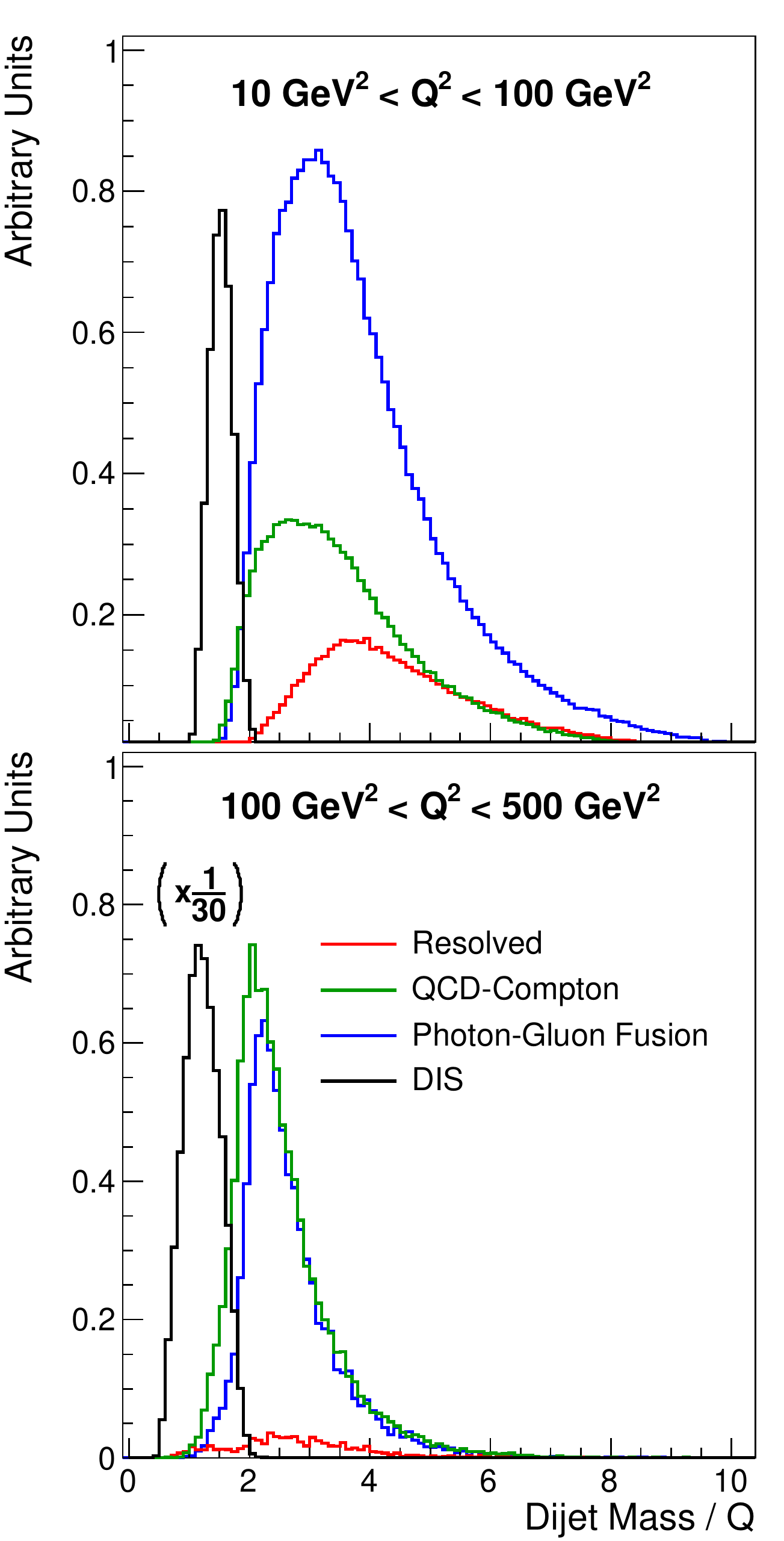}
	\caption[Cut to Remove DIS Events]{[color online] Ratio of dijet invariant mass over $Q$ for the resolved, QCDC, PGF, and DIS subprocesses in \QTWO bins of 10-100~GeV$^2$ (upper panel), and 100-500~GeV$^2$ (lower panel). This ratio allows the separation of dijet events arising from the leading order DIS subprocess from all others. Note that each panel has been scaled an arbitrary amount and in the lower panel, the DIS curve has been scaled down by an additional factor of 30.}
	\label{fig:DISCUT}
\end{figure}

As they are largely a low \QTWO phenomenon, a significant fraction of resolved events can be 
eliminated simply by requiring that $\QTWO > 1$~GeV$^2$. However, because requiring two high-\PT jets 
significantly biases the event sample against leading order DIS and toward higher-order 
processes, a non-negligible resolved contribution remains for $\QTWO > 1$~GeV$^2$ (see Fig. \ref{fig:RECOXGAMMA}).
This remaining resolved component can be greatly 
reduced by requiring that the reconstructed $x_{\gamma}$ be close to unity. As explained 
in \cite{Chu:2017mnm}, the virtual photon behaves as a point particle for the PGF and QCDC 
processes, meaning it contributes 100\% of its momentum to the interaction and thus should have 
$x_{\gamma} = 1$. Conversely, for a resolved event, the photon behaves as a composite 
particle and only a fraction of its momentum contributes to the interaction, meaning $x_{\gamma}$ 
will have a broad distribution of values less than unity. Figure \ref{fig:RECOXGAMMA} presents 
the reconstructed $x_{\gamma}$ distributions for dijets from the resolved, QCDC, PGF, 
and leading order DIS 
processes for three 
$Q^2$ ranges. It is clear that a cut on $x_\gamma$ can effectively remove a large fraction of the resolved 
contribution while preserving most of the direct events. Cuts on $x_\gamma$ of 
0.75 and 0.60 were placed for the \QTWO ranges $1-10$~GeV$^2$ and $10-100$~GeV$^2$,  
respectively, which reduce the resolved component to less than 10\% of the remaining PGF contribution. No cut 
is placed on the 100-500~GeV$^2$ bin. It should be noted that the dijets from leading 
order DIS which appear at larger \QTWO values arise when the target remnant receives a large enough transverse 
kick to form a jet which passes the selection criteria. Such events can be eliminated, with minimal loss to 
PGF and QCDC yields, by cutting on the ratio of dijet mass to $Q$ (the ratio was required to be greater than 2.0) 
as shown 
in Fig. \ref{fig:DISCUT}. For the following, residual contributions from resolved and leading order DIS events were 
omitted for simplicity as neither subprocess was found to contribute significantly to the expected asymmetry.

\begin{figure*}[t!]
	\centering \includegraphics[keepaspectratio=true, width=0.95\linewidth]{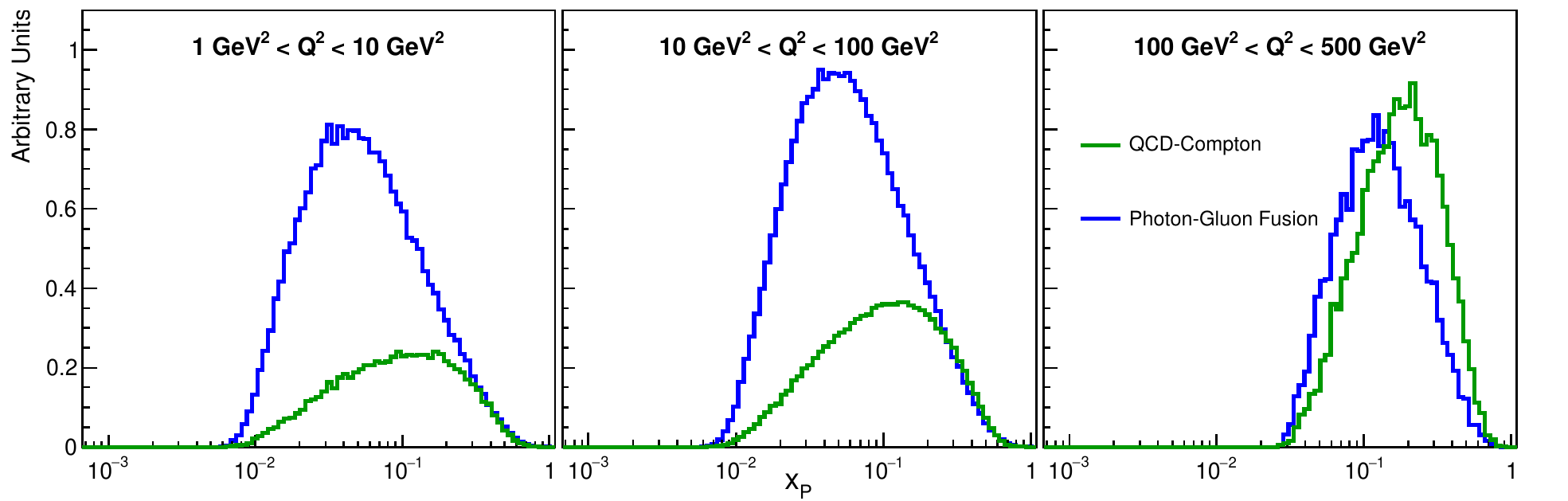}
	\caption[Reconstructed $x_{P}$]{[color online] Reconstructed momentum fraction of the parton arising from the proton for the QCDC and PGF subprocesses in \QTWO bins of 1-10~GeV$^2$ (left), 10-100~GeV$^2$ (middle), and 100-500~GeV$^2$ (right).}
	\label{fig:XPROTON}
\end{figure*}

\begin{figure}[t!]
	\centering \includegraphics[keepaspectratio=true, width=0.95\linewidth]{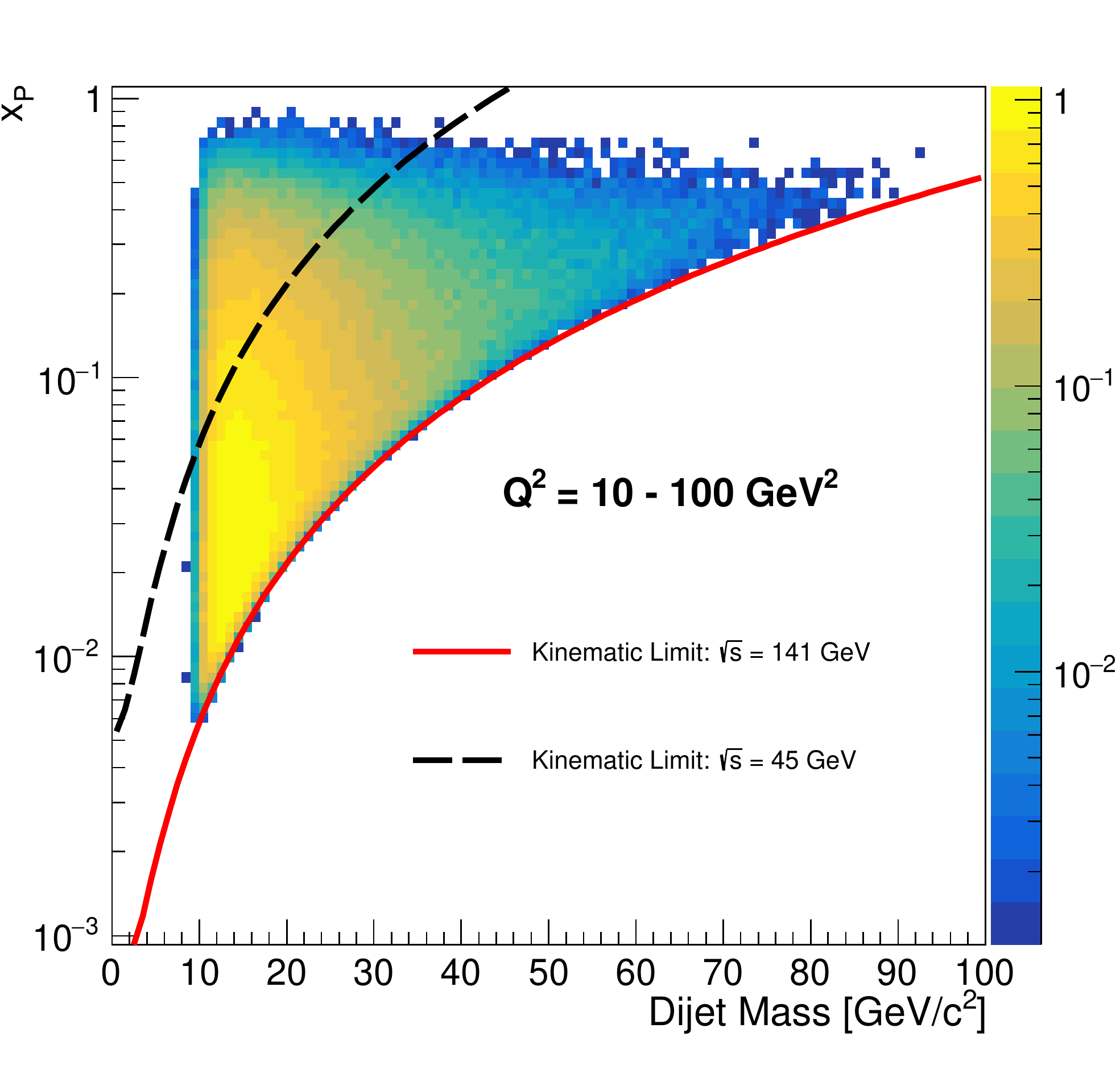}
	\caption[$x_{P} Vs Mass$]{[color online] Reconstructed momentum fraction of the parton arising from the proton vs the invariant mass of the resulting dijet for the QCDC and PGF processes combined and \QTWO between 10 and 100~GeV$^2$. The solid red and dashed black lines denote the allowed phasespace for $\sqrt{s} = 141$ and 45~GeV, respectively.}
	\label{fig:XPVSMASS}
\end{figure}

Removing the QCDC contribution is not as straightforward as eliminating resolved events because the 
PGF and QCDC processes have very similar event topologies. However, as can be seen in 
Fig. \ref{fig:XPROTON}, the PGF cross section peaks at lower values of $x_{p}$ relative to QCDC events. 
Thus, at least at \QTWO below 100~GeV$^2$ where the the QCDC cross section is small compared to PGF, 
$x_{p}$ can be used 
to select regions of high or low signal-to-background. It should be noted that placing an upper $x_{p}$ cut will 
restrict the maximum accessible dijet mass as seen in Fig. \ref{fig:XPVSMASS}. 
The relationship between $x_{p}$ and dijet mass (at leading order) is given by the Eq. \ref{eq:XPMASS}:

\begin{equation} \label{eq:XPMASS}
x_{p} = x_{\mathrm{B}} + \frac{M_{jj}^{2}}{sy},
\end{equation}

\noindent where $x_{\mathrm{B}}$ is Bjorken-$x$, $M_{jj}$ is the invariant mass of the dijet system, 
$s$ is the center-of-mass energy, and $y$ is the inelasticity. Equation \ref{eq:XPMASS} shows that 
the $x_{P}$ values accessible to this measurement are driven by the center-of-mass energy and that 
for a minimum dijet mass of 10~GeV$^2$, $s$ of 20000~GeV$^2$, and a maximum inelasticity of 0.95, 
the lowest $x_{p}$ available is roughly $5 \times 10^{-3}$. Figure \ref{fig:XPVSMASS} presents  
accessible $x_{P}$ values as a function of dijet invariant mass as well as curves delineating the 
available phase-space for center-of-mass energies of 141 and 45~GeV.

\subsection{Expected Asymmetry}

The method used here to determine the behavior of $A_{LL}$ is the same as in \cite{Chu:2017mnm}, 
which was adapted from \cite{Airapetian:2010ac}. For each simulated event, a weight was calculated using the 
subprocess and kinematic information from PYTHIA as well as external (un)polarized PDFs. 
The asymmetry is then found as the average over these weights. The weights are calculated according to: 

\begin{equation} \label{eq:ASYMWEIGHT}
	w = \hat{a}(\hat{s},\hat{t},\mu^2,Q^2) \frac{\Delta f^{\gamma^{*}}_{a}(x_a,\mu^2)}{f^{\gamma^{*}}_{a}(x_a,\mu^2)} \frac{\Delta f^{N}_{b}(x_b,\mu^2)}{f^{N}_{b}(x_b,\mu^2)},
\end{equation}

\noindent where $\hat{a}(\hat{s},\hat{t},\mu^2,Q^2)$ is the subprocess dependent parton 
level asymmetry, 
$(\Delta) f^{\gamma^{*}}_{a}(x_a,\mu^2)$ is the (polarized) PDF for the virtual photon, 
and $(\Delta) f^{N}_{b}(x_b,\mu^2)$ is the (polarized) PDF for the proton. 
The leading order expressions for $\hat{a}$ were taken from \cite{Airapetian:2010ac} and include the appropriate 
depolarization factors. The DSSV14 \cite{deFlorian:2014yva} and NNPDFpol1.1 \cite{Nocera:2014gqa} sets were used to describe the 
polarized proton and were normalized by the MSTW2008 \cite{Martin:2009iq} and NNPDF2.3 \cite{Ball:2012cx} unpolarized PDFs, 
respectively. Because only direct events were considered, the $\Delta f^{N}_{b}(x_b,\mu^2)/f^{N}_{b}(x_b,\mu^2)$ 
term is identically unity.

\begin{figure}[t!]
	\centering \includegraphics[keepaspectratio=true, width=0.95\linewidth]{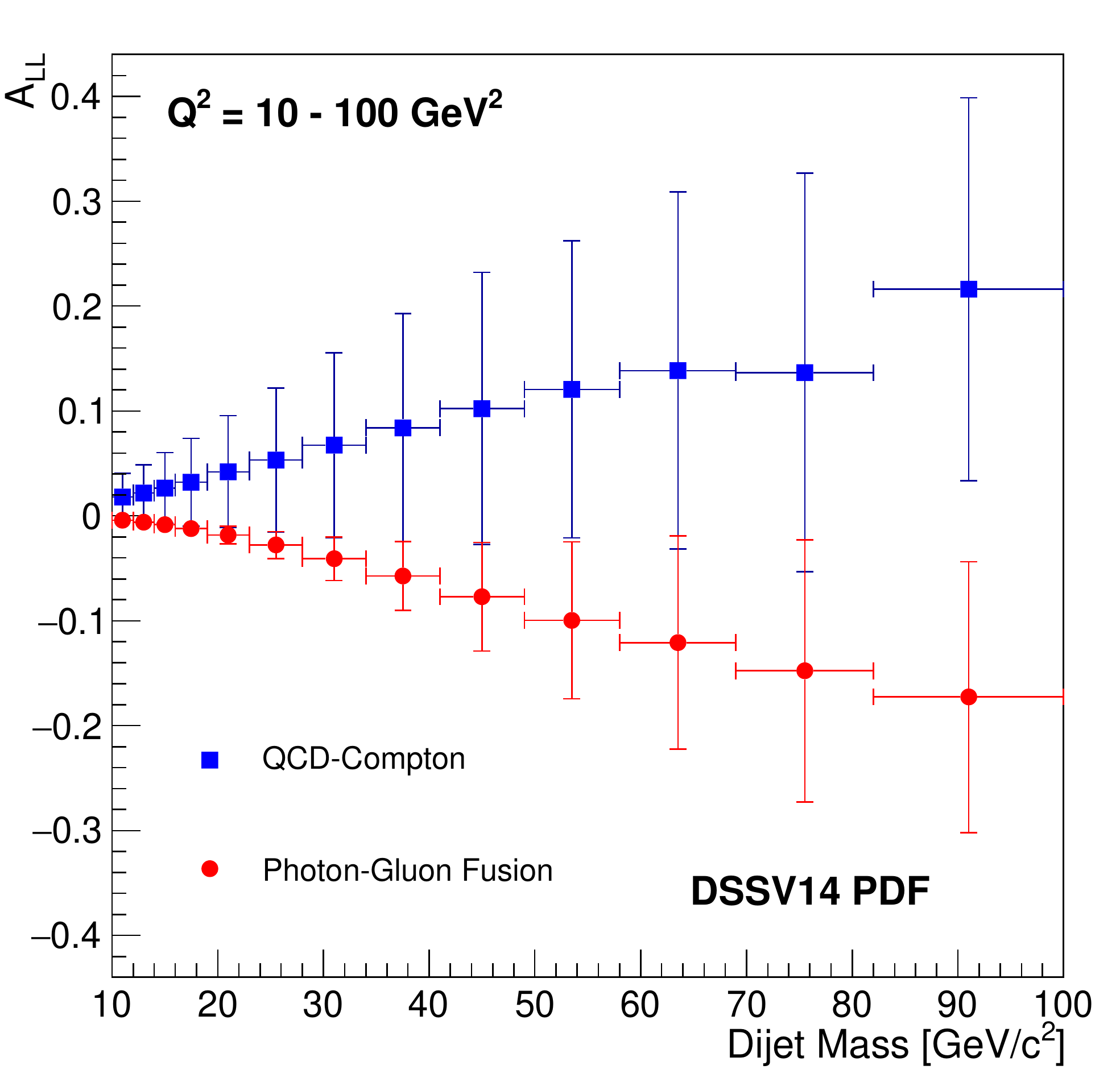}
	\caption[$A_{LL}$ vs Mass by Subprocess]{[color online] Dijet $A_{LL}$ as a function of dijet invariant mass for the QCDC and PGF subprocesses in the 10-100~GeV$^{2}$ \QTWO bin}
	\label{fig:ALLSUB}
\end{figure}

Figure \ref{fig:ALLSUB} shows $A_{LL}$ as a function of dijet invariant mass for the QCDC and PGF subprocesses 
obtained using the DSSV set (NNPDF is similar) for \QTWO between 10 and 100~GeV$^2$. The error bars represent the 
RMS of the distribution of weights in 
each mass bin. The width of the weight distribution is larger for the QCDC process because the sign of the weight 
can change due to different asymmetry signs for up and down quarks. For the PGF process on the other hand, the 
gluon asymmetry is positive everywhere in the relevant kinematics and the $\hat{a}$ term is always negative, meaning 
the weight is always negative and therefore the spread in weights is smaller. 
It is seen that the asymmetries for each subprocess grow with dijet mass and become sizable, 
reaching values of 20\% for the highest masses. However, because the QCDC and PGF asymmetries are roughly equal 
in magnitude but opposite in sign, one can expect that the total asymmetry will be significantly smaller than 
the asymmetry for either individual subprocess.

\begin{figure}[t!]
	\centering \includegraphics[keepaspectratio=true, width=0.95\linewidth]{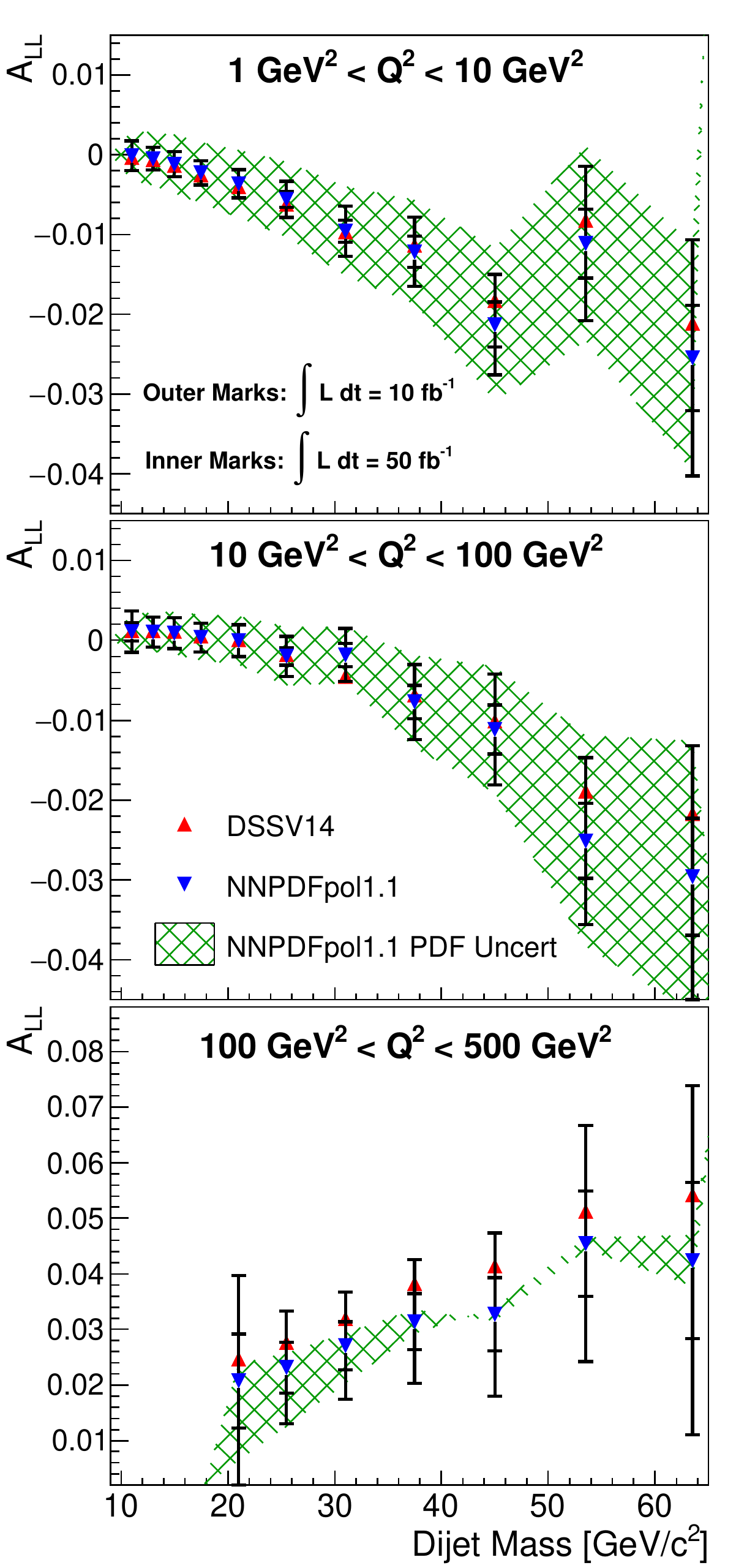}
	\caption[Dijet $A_{LL}$]{[color online] Dijet $A_{LL}$ as a function of dijet invariant mass for the combined QCDC and PGF subprocesses using the DSSV14 and NNPDF1.1 polarized PDFs in the 1-10~GeV$^{2}$ (top), 10-100~GeV$^{2}$ (middle), and 100-500~GeV$^{2}$ (bottom) \QTWO bins. Note that projected statistical uncertainties for the DSSV14 points are not shown for clarity, but are nearly identical to those from NNPDF1.1.}
	\label{fig:ALLTot}
\end{figure}

The combined QCDC and PGF \ALL obtained using both the DSSV and NNPDF PDFs can be seen in Fig \ref{fig:ALLTot} as a function of 
dijet invariant mass in three \QTWO ranges. The error bars represent expected statistical uncertainties calculated according 
to:

\begin{equation}
\sigma = \frac{1}{\mathrm{P_{e}P_{p}}}\sqrt{\frac{1}{\mathrm{N}} - \frac{A_{LL}^2}{\mathrm{N}}},
\end{equation}

\noindent where $\mathrm{P_{e}}$ and $\mathrm{P_{p}}$ are the electron and proton beam polarizations 
(taken as 80\% and 70\%, respectively) and N is the number of expected events assuming 
an integrated luminosity of 10~fb$^{-1}$ or 50~fb$^{-1}$. The green bands represent the uncertainty on 
the NNPDFpol1.1 polarized PDF. As expected, the QCDC and PGF asymmetries cancel to a large degree, resulting in maximum asymmetries 
of a few percent.

\begin{figure*}[t!]
	\centering \includegraphics[keepaspectratio=true, width=0.95\linewidth]{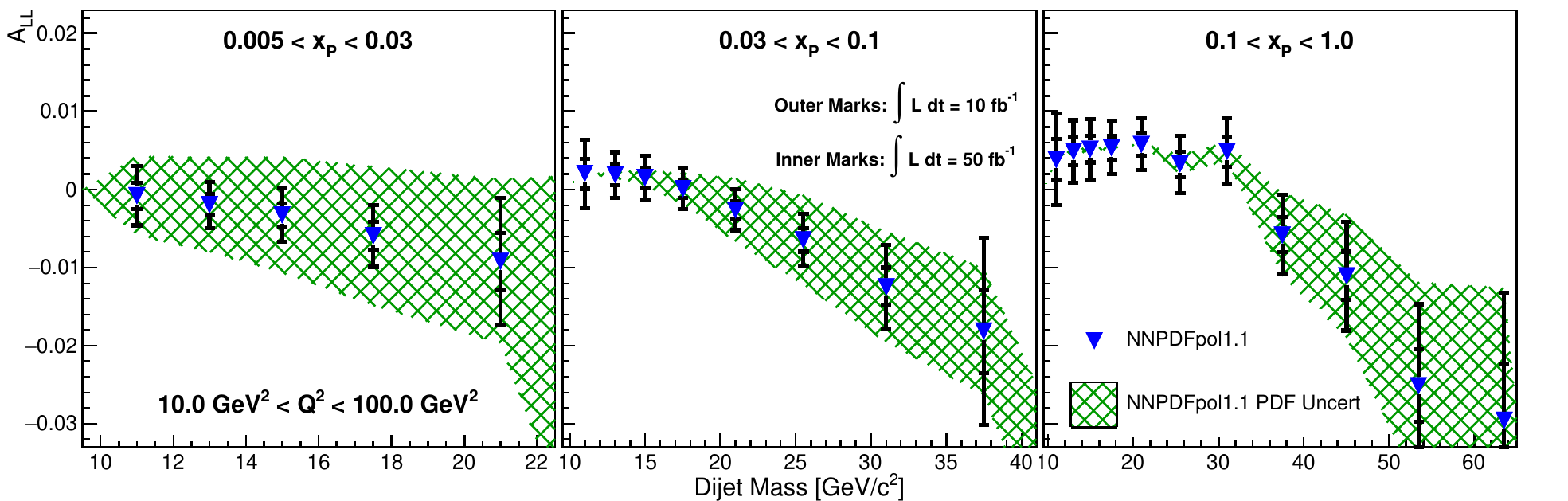}
	\caption[Dijet $A_{LL}$ $x_{P}$]{[color online] Dijet $A_{LL}$ as a function of dijet invariant mass for the combined QCDC and PGF subprocesses using the NNPDF1.1 polarized PDFs in the 10-100~GeV$^{2}$ \QTWO bin for partonic momentum fractions of $0.005 < x_{P} < 0.03$ (left), $0.03 < x_{P} < 0.1$ (middle), and $0.1 < x_{P} < 1.0$ (right).}
	\label{fig:ALLXP}
\end{figure*}

In order to better isolate the gluon contribution, it would be helpful to reduce the fraction 
of QCDC events, which carry information on the quarks.
As mentioned above, the reconstructed momentum fraction carried by the parton from the proton 
can be used to select kinematic regions where the PGF subprocess is dominant. 
Figure \ref{fig:ALLXP} presents dijet $A_{LL}$ as a function of invariant mass for the bin  
$Q^2 = 10-100$~GeV$^2$ for three 
$x_{p}$ slices: $0.005 < x_{p} < 0.03$, $0.03 < x_{p} < 0.1$, and $0.1 < x_{p} < 1.0$, with the 
ratio of PGF to QCDC events decreasing with increasing $x_{p}$. Note that only NNPDF1.1 results are 
shown for clarity. The bars again show expected 
statistical uncertainties assuming 10 and 50~fb$^{-1}$ and the green bands are the uncertainty on 
the NNPDFpol1.1 polarized PDFs. The effects of slicing in $x_{p}$ are modest, but do shift the asymmetries 
to more negative values and increase the ratio of PDF to statistical uncertainties.

The PDF uncertainties presented in Fig. \ref{fig:ALLTot} and \ref{fig:ALLXP} represent the current 
state of knowledge 
on the helicity structure of the proton. These uncertainties will shrink substantially with the 
addition of inclusive $g_{1}$ measurements, which will be the golden channel for the constraint of 
$\Delta g(x,Q^2)$. Figures \ref{fig:ALLTot} and \ref{fig:ALLXP} show that substantial integrated 
luminosities will be needed in order for the dijet measurements to improve on our current knowledge of 
$\Delta g(x,Q^2)$ meaning it will be unlikely the dijet measurement can compete directly with $g_1$ in 
constraining the gluon contribution to the proton spin. The benefit of the dijet measurement will likely 
be in its complementarity to $g_1$ as the dijets arise from different subprocesses and will have different 
associated systematics than inclusive observables. 

\section{Summary and Outlook} \label{sec:SUMMARY}

Jet observables have proven their utility as probes of the subatomic realm at virtually all high energy colliders 
operated to date, while recent experimental and theoretical advances, spurred by the success of modern colliders such 
as the LHC, have seen jets become true precision probes. This success behooves those interested in the science 
an EIC will address to explore the potential benefits that jets could provide. To that end, this paper has 
systematically explored a number of topics relevant to the experimental analysis of jets at an EIC. 

The first issues addressed were particulars of the actual jet finding. There was no significant dependence seen on the 
choice of jet algorithm, but jets with larger radii were found to better reproduce the underlying partonic 
kinematics and the anti-k$_\mathrm{T}$ algorithm with $R = 1.0$ was chosen for all subsequent studies. Next, jet 
kinematic distributions were quantified, comparing inclusive jet $p_\mathrm{T}$, dijet mass, and pseudorapidity 
spectra for both inclusive jets and dijets for a range of \QTWO values, subprocesses, and center-of-mass energies. 
It was seen that higher center-of-mass energies produce greater yields of jets/dijets, especially at larger 
$p_\mathrm{T}$/mass. The pseudorapidity of jets was also seen to increase with $\sqrt{s}$, driven by the larger boost 
imparted by higher hadron beam energy. The energy 
contribution from underlying event activity was also studied using two different methods and was found to be small, 
although it will need to be considered when dealing with low \PT jets, where even small underlying event contributions 
can have a fractionally larger effect. Distortions 
of the jet \PT due to realistic detector resolutions were investigated using a smearing program tuned to replicate the BeAST 
detector design and were found to be minor. Special attention was given to the role of hadron calorimetry at mid-rapidity 
with high resolution, low resolution and no calorimeter options explored. A scheme to use a low resolution calorimeter as 
a neutral hadron veto system with the goal of improving the overall jet energy resolution was also discussed. 
Finally, an example analysis was presented in which dijets were used to tag 
photon-gluon fusion events for the purpose of constraining the gluon helicity contribution to the proton spin. Methods for 
reducing background and isolating PGF events were demonstrated and the expected asymmetries and their uncertainties were 
found and compared to current knowledge of gluon polarizations.

While this paper provides a solid introduction to the experimental reality of jet physics at an EIC, the topic is still 
relatively new and more detailed follow-up studies will be needed to build a robust EIC jet program. Areas of future study 
include potentially fruitful 
topics such as jet substructure and the use of jets in \EA collisions, which were not addressed here at all. In addition, 
more realistic detector simulation and modeling will be needed in order to inform detector performance requirements. 
We hope this paper will serve as a valuable resource and jumping off point for both theorists and experimentalists who wish 
to further pursue jet topics at the EIC.

\begin{acknowledgments}
We would like to thank Felix Ringer, Kyle Lee, Kolja Kauder, Miguel Arratia, Barbara Jacak, and 
Frank Petriello for helpful discussions. We are grateful to Alexander Kiselev for 
providing details on proposed tracking resolutions for the BeAST detector design. B.P. is supported 
by the Program Development program at Brookhaven National Laboratory, while E.C.A and X.C. 
acknowledge support from the U.S. Department of Energy under contract number de-sc0012704.
\end{acknowledgments}

\clearpage
\bibliography{jetPaperBib}

\end{document}